\documentclass[fleqn,usenatbib]{mnras}

\usepackage[T1]{fontenc}
\usepackage{ae,aecompl}

\usepackage{graphicx}	
\usepackage{amsmath}	
\usepackage{amssymb}	

\usepackage{xcolor}

\title{EPIC 219217635: A Doubly Eclipsing Quadruple System Containing an Evolved Binary}

\author[Borkovits et al.]{
T.~Borkovits$^{1,2}$,
S.~Albrecht$^3$,
S.~Rappaport$^4$,   
L.~Nelson$^5$,
A.~Vanderburg$^{6,7,8}$, 
\newauthor
B.L.~Gary$^9$, 
T.G.~Tan$^{10}$,
A.B.~Justesen$^{11}$,
M.H.~Kristiansen$^{12,13}$,
T.L.~Jacobs$^{14}$,
\newauthor
D.~LaCourse$^{15}$,
H.~Ngo$^{16,17}$,
N.~Wallack$^{17}$,
G.~Ruane$^{18}$,
D.~Mawet$^{19}$,
S.B.~Howell$^{20}$, 
\newauthor
R.~Tronsgaard$^{21,22}$
\\ \\
$^{1}$ Baja Astronomical Observatory of Szeged University, H-6500 Baja, Szegedi \'{u}t, Kt. 766, Hungary; borko@electra.bajaobs.hu \\
$^{2}$ Konkoly Observatory, Research Centre for Astronomy and Earth Sciences, Hungarian Academy of Sciences, \\ Konkoly Thege Miklós \'ut 15-17, H-1121 Budapest, Hungary \\
$^{3}$ Stellar Astrophysics Centre, Department of Physics and Astronomy Aarhus University, Ny Munkegade 120, 8000, Aarhus C, Denmark \\
$^{4}$ Department of Physics, and Kavli Institute for Astrophysics and Space Research, M.I.T., Cambridge, MA 02139, USA; sar@mit.edu \\
$^{5}$ Department of Physics and Astronomy, Bishop's University, 2600 College St., Sherbrooke, QC J1M 1Z7; lnelson@ubishops.ca \\ 
$^{6}$ Harvard-Smithsonian Center for Astrophysics, 60 Garden Street, Cambridge, MA 02138 USA; avanderburg@cfa.harvard.edu \\ 
$^{7}$ Department of Astronomy, The University of Texas at Austin, 2515 Speedway, Stop C1400, Austin, TX 78712 \\
$^{8}$ NASA Sagan Fellow \\
$^{9}$ Hereford Arizona Observatory, Hereford, AZ 85615; BLGary@umich.edu \\
$^{10}$ Perth Exoplanet Survey Telescope, Perth, Western Australia 6010 \\
$^{11}$  Stellar Astrophysics Centre, Department of Physics and Astronomy, Aarhus University, Ny Munkegade 120, DK-8000 Aarhus C, Denmark\\
$^{12}$ DTU Space, National Space Institute, Technical University of Denmark, Elektrovej 327, DK-2800 Lyngby, Denmark \\
$^{13}$ Brorfelde Observatory, Observator Gyldenkernes Vej 7, DK-4340 T\o ll\o se, Denmark \\
$^{14}$ 12812 SE 69th Place Bellevue, WA 98006 \\
$^{15}$ 7507 52nd Place NE Marysville, WA 98270 \\
$^{16}$  NRC Canada Herzberg Astronomy \& Astrophysics, Victoria, BC, Canada \\
$^{17}$ Division of Geological and Planetary Sciences, California Institute of Technology, Pasadena, CA, USA \\
$^{18}$ Department of Astronomy, California Institute of Technology, Pasadena, CA, USA and NSF Astronomy and Astrophysics Postdoctoral Fellow \\
$^{19}$ Department of Astronomy and Jet Propulsion Laboratory, California Institute of Technology, Pasadena, CA, USA \\
$^{20}$ Space Science \& Astrobiology Division, NASA Ames Research Center, M/S 245-1, Moffett Field, CA 94035 \\ 
$^{21}$ DTU Space, National Space Institute, Technical University of Denmark, Elektrovej 328, DK-2800 Kgs. Lyngby, Denmark.\\
$^{22}$ Nordic Optical Telescope, Rambla Jos\'e Ana Fern\'andez P\'erez 7, 38711 Bre\~na Baja, Spain \\
}

\date{}

\pubyear{}

\begin{document}
\label{firstpage}
\pagerange{\pageref{firstpage}--\pageref{lastpage}}
\maketitle

\begin{abstract}
We have discovered a doubly eclipsing, bound, quadruple star system in the field of K2 Campaign 7.  EPIC 219217635 is a stellar image with $Kp = 12.7$ that contains an eclipsing binary (`EB') with $P_A = 3.59470$ d and a second EB with $P_B = 0.61825$ d.   We have obtained followup radial-velocity (`RV') spectroscopy observations, adaptive optics imaging, as well as ground-based photometric observations.   From our analysis of all the observations, we derive good estimates for a number of the system parameters.  We conclude that (1) both binaries are bound in a quadruple star system; (2) a linear trend to the RV curve of binary A is found over a 2-year interval, corresponding to an acceleration, $\dot \gamma = 0.0024 \pm 0.0007$ cm s$^{-2}$; (3) small irregular variations are seen in the eclipse-timing variations (`ETVs') detected over the same interval;  (4) the orbital separation of the quadruple system is probably in the range of 8-25 AU; and (5) the orbital planes of the two binaries must be inclined with respect to each other by at least 25$^\circ$.  In addition, we find that binary B is evolved, and the cooler and currently less massive star has transferred much of its envelope to the currently more massive star. We have also demonstrated that the system is sufficiently bright that the eclipses can be followed using small ground-based telescopes, and that this system may be profitably studied over the next decade when the outer orbit of the quadruple is expected to manifest itself in the ETV and/or RV curves. 
\end{abstract}




\section{Introduction}

Quadruple, or higher-order multiple systems constitute a relatively small, but very important fraction of gravitationally bound, few-body stellar systems. For example, according to the distance limited ($D\leq75$\,pc) sample of \citet{derosaetal14} the lower limit on the frequency of quadruple or higher-order multiple systems\footnote{In these surveys single A, F, or G type stars are to be counted as `systems'.} having an A-type star as the more massive component is about 2.5\%. Investigating a similar distance-limited ($D\leq67$\,pc) collection of FG dwarf multiples \citet{Tokovinin14} found the same occurrence frequency to be 4\%. The majority of the known quadruple stars form a 2+2 hierarchy, i.e. two smaller-separation (and, therefore, shorter-period) binaries which orbit around their common centre of mass on a much wider, longer-period orbit. For example, in the previously mentioned sample of FG multiples, 37 of the 55 quadruple stars have the 2+2, double-binary configuration. Furthermore, quadruple subsystems of higher-order multiple-star systems also often come in the form of a 2+2 hierarchy.

Double binary systems are important tracers of stellar formation scenarios. Their mass and period ratios, as well as their flatness (i.e. the inclination of the outer orbit relative to the two inner ones) may carry important information on their formation processes, as well as their further evolution \citep[see, e.g.][and references therein]{tokovinin08,tokovinin18}.

Another interesting aspect of double binaries is their dynamics, i.e. long-term orbital evolution. Recent analytical \citep{fangetal18} as well as numerical \citep{pejchaetal13} studies have pointed out that 2+2 quadruples with an inclined outer orbit may be subject to Kozai-Lidov-cycles \citep{kozai62,lidov62} that reach higher eccentricities than triple stars.  This can result, amongst other interesting phenomena, in dramatic inner-binary eccentricity oscillations which temporarily might produce extremely high eccentricities (such as, e.g. $e_\mathrm{in}\geq0.999$) for a remarkable fraction of the possible 2+2 quadruple systems. In turn, this may lead to stellar mergers, thereby forming hierarchical triples or producing blue stragglers \citep{peretsfabrycky09}, not to mention the possibility of the merger of two white dwarfs, producing a type Ia SN explosion (see the short summary regarding this question in \citealt{fangetal18}). Furthermore, a less extreme scenario can also be the formation of tight binaries \citep[see, e.g.][]{eggletonkiseleva-eggleton01,fabryckytremaine07,naozfabrycky14}.   

Doubly eclipsing quadruples constitute a remarkable subclass of 2+2 quadruple systems (and/or subsystems), where both inner binaries exhibit eclipses. The first known, and for some decades the sole representative, of these objects is the pair of W UMa-type eclipsing binaries  (`EBs') BV and BW Dra \citep{battenhardie65}. The discovery of the second member of this group (V994 Her) was reported more than four decades later \citep{leeetal08}. During the last decade, however, due to the advent of the long-duration, almost continuous photometric sky-surveys, both ground-based (e.g. SuperWASP, \citealt{SWASP}, OGLE, \citealt{OGLE}, etc.) and space photometry (especially {\em Kepler}, \citealt{Borucki}, and CoRoT space telescopes, \citealt{CoRoT}), several new doubly eclipsing quadruple {\em candidates} have been discovered photometrically. Some examples, without any attempt at completeness, are KIC\,4247791 \citep{Lehmann12}, CzeV343 \citep{Cagas}, 1SWASP\,J093010.78+533859.5 \citep{Lohr}, EPICs\, 212651213 \citep{Rappaport16} and 220204960 \citep{Rappaport17}. (Some of these quadruples have farther, more distant, and also likely bound companions as well.) Another, extraordinarily interesting system is KIC\,4150611, which consists of three or four eclipsing binaries, and one ``binary'' of the double binary configuration is itself a triply eclipsing triple subsystem \citep{TheThing,helminiaketal17}.   Additional blended EB lightcurves amongst CoRoT and {\em Kepler} targets were reported by \citet{eriksonetal12,fernandezchou15,hajduetal17} and \citet{Borko16}. 

One should note, however, that by observing only a lightcurve which is characterized by the blended light of two EBs, one cannot be certain that the two EBs really form a gravitationally bound system. The small separation or even the unresolved nature of the optical images of the sources, as well as reasonably similar radial velocities and/or proper motions can be very good indirect indicators of the bound nature of the pairs, but definitive evidence can be obtained only if the relative motion, or any other dynamical interactions of the two binaries, can be observed. Regarding these latter strict requirements, at this moment, to the best of our knowledge, there are only three pairs of EBs exhibiting blended lightcurves, for which their gravitationally bound, quadruple nature is beyond doubt. These are V994 Her \citep{V994Her}, V482 Per \citep{torresetal17} in which cases the light travel-time effect (LTTE) was clearly detected, and EPIC\,220204960 \citep{Rappaport17} which exhibits dynamically forced rapid apsidal motions in both binaries.\footnote{Most recently \citet{hongetal18} have published an analysis of two double EB candidates in the Large Magellanic Cloud, namely OGLE-LMC-ECL-15674 and OGLE-LMC-ECL-22159. The binaries in the first system exhibit rapid eclipse depth variations, and therefore, probably inclination variations, and one of them also shows rapid apsidal motion. Thus, with high likelihood, this object is also a dynamically interactive, bound quadruple system.}

In this work we report the discovery with NASA's {\em Kepler} space telescope during campaign 7 of its two-wheeled mission (hereafter referred to as `K2') of a quite likely physically bound quadruple system consisting of two eclipsing binaries, with orbital periods of 3.59470 d and 0.61825 d. We derive many of the parameters for this system. The paper is organized as follows. In Sect.~\ref{sec:K2} we describe the 80-day K2 observation of EPIC\,219217635 with its two physically associated eclipsing binaries. 
We have obtained Keck AO imaging of the target star (see Sect.~\ref{sec:Keck_AO}), and we find that the two binaries are unresolved down to $\sim$$0.05''$. In Sect.~\ref{sec:photometry} we discuss the eight eclipse minima that we were able to measure with ground-based photometry and analyse them together with the other eclipse minima determined from the 80-day-long K2 lightcurve in Sect.\,\ref{sec:ETV}. We obtained 20 radial-velocity spectra which lead to mass functions for the two binaries; these are described in Sect.~\ref{sec:NOT-FIES}. We then use our improved lightcurve and RV curve emulator to model and evaluate both the eclipsing binary lightcurves and the RV curves simultaneously (see Sect.~\ref{sec:model}). In Sect.~\ref{sec:outerorbit} we explore the constraints we can place on the parameters of the outer quadruple orbit. In Sect.~\ref{sec:evolution} we investigate the likely mass-transfer evolution that has occurred in binary B.  Finally, we summarize our findings and draw some conclusions in Sect.~\ref{sec:conclusion}.

\vspace{0.6cm}


\section{K2 Observations}
\label{sec:K2}

As part of our ongoing search for eclipsing binaries, we downloaded all available K2 Extracted Lightcurves common to Campaign 7 from the Mikulski Archive for Space Telescopes (`MAST')\footnote{\url{http://archive.stsci.edu/k2/data\_search/search.php}}. We utilized both the Ames pipelined data set and that of \citet{AV}.  The flux data from all 24,000 targets were searched for periodicities via Fourier transforms and the BLS algorithm \citep{Kovacs}. The folded lightcurves of targets with significant peaks in their FFTs or BLS transforms were then examined by eye to look for unusual objects among those with periodic features.  In addition, some of us (MHK, DL, and TLJ) visually inspected all the K2 lightcurves for unusual stellar or planetary systems with {\tt LcTools} (Kipping et al.~2015).  

\begin{figure*}
\begin{center}
\includegraphics[width=0.9 \textwidth]{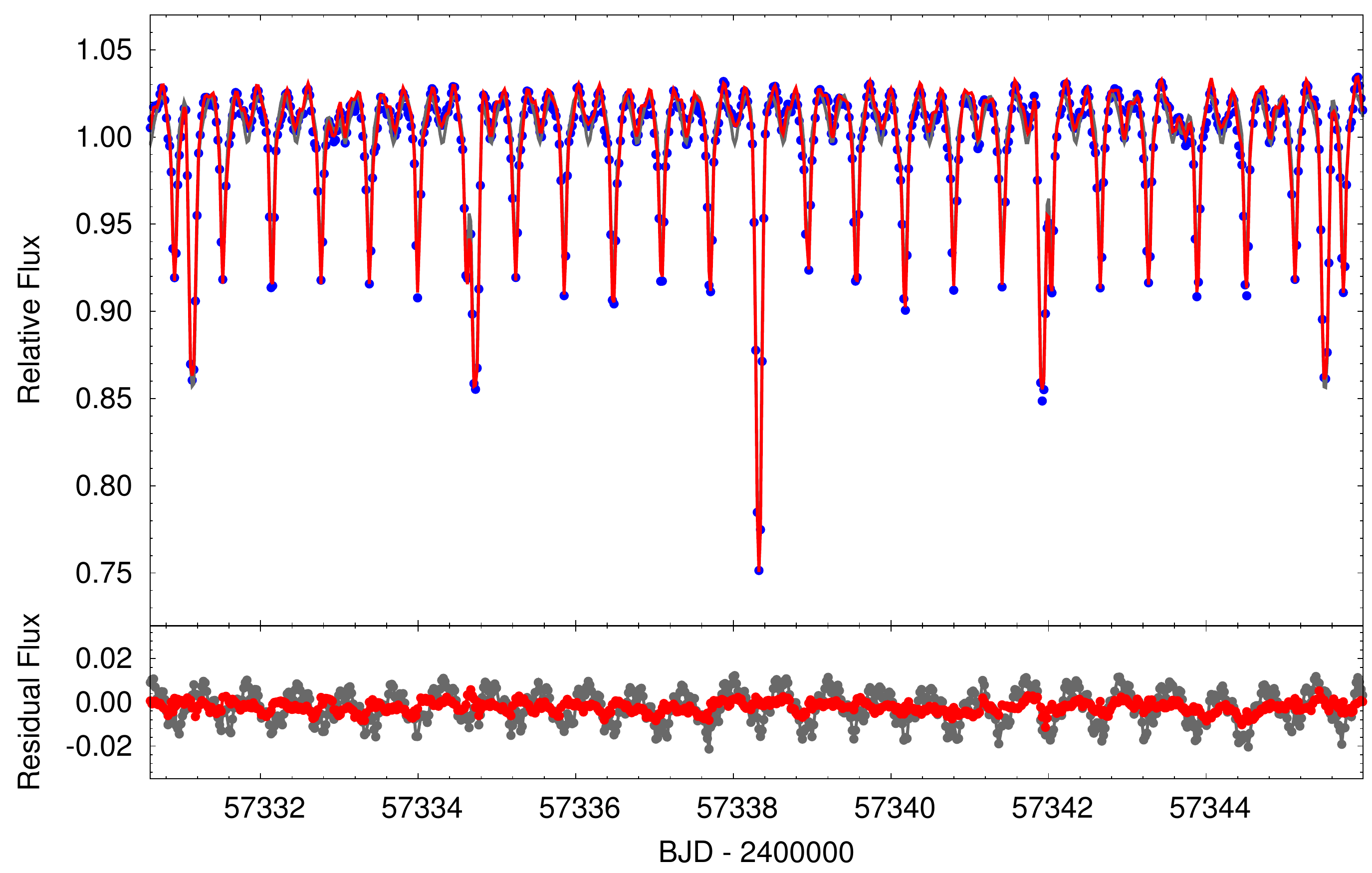}
\caption{A zoomed-in $\sim14$-day segment of the K2 flux data showing the superposition of the eclipses of the A and B binaries.  The data are shown in blue, the gray curve is a pure, double, blended eclipsing binary model fit, while the red curve is the net model fit taking into account both the binary and the other distortion effects (see text for details).  The residuals of the data from the two models are shown in the bottom panel.}
\label{fig:zoomLC}
\end{center}
\end{figure*}

Within a day after the release of the Field 7 data set, EPIC 219217635 was identified as a potential quadruple star system by both visual inspection and via the BLS algorithmic search. A two-week-long section of the K2 lightcurve is shown in Fig.~\ref{fig:zoomLC}, where several features can be seen by inspection. The eclipses of the 3.595-day `A' binary and 0.618-day `B' binary are fairly obvious. Each binary has a deep and a shallow eclipse.     

\begin{figure}
\begin{center}
\includegraphics[width=0.48 \textwidth]{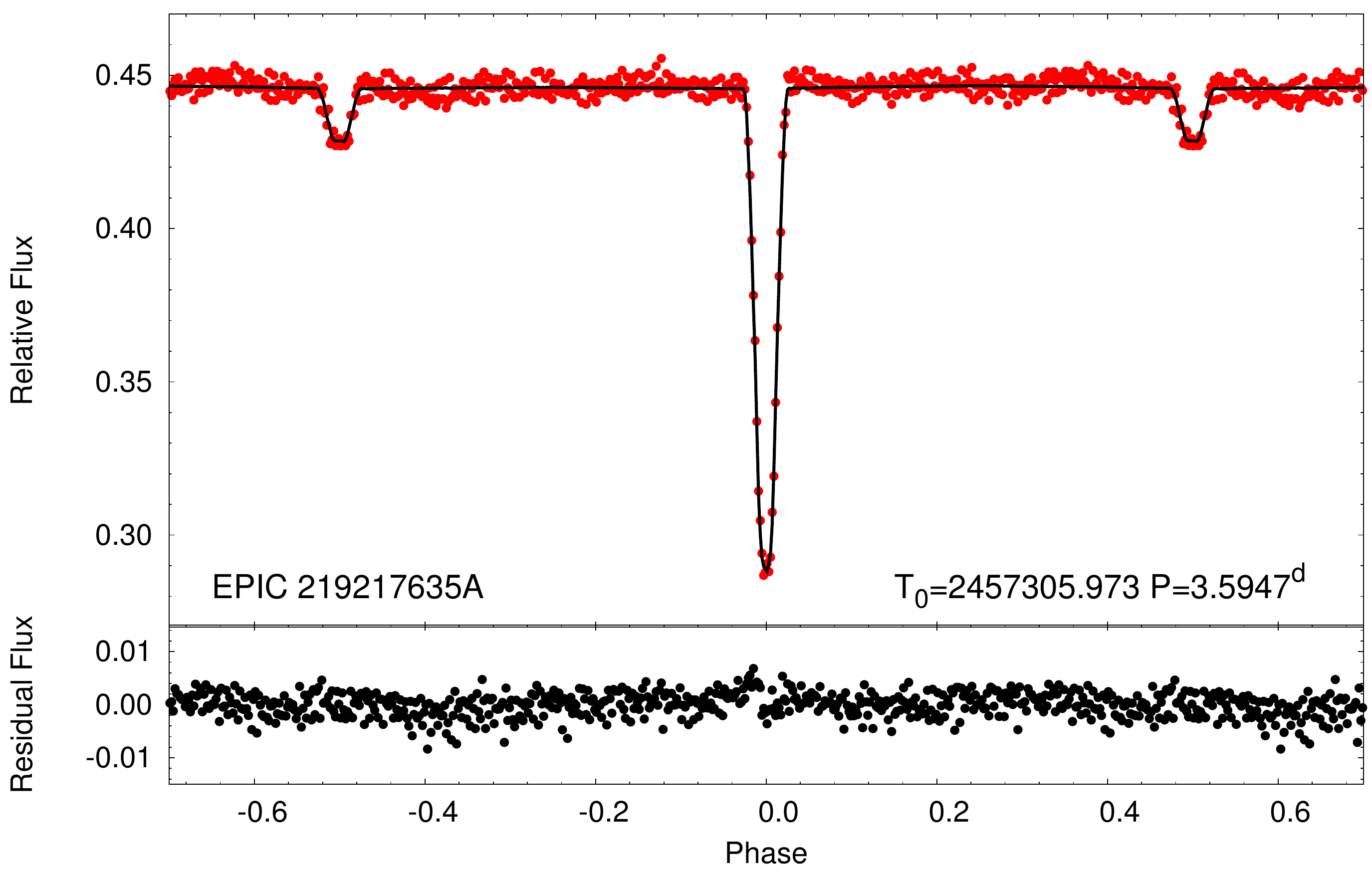}
\includegraphics[width=0.48 \textwidth]{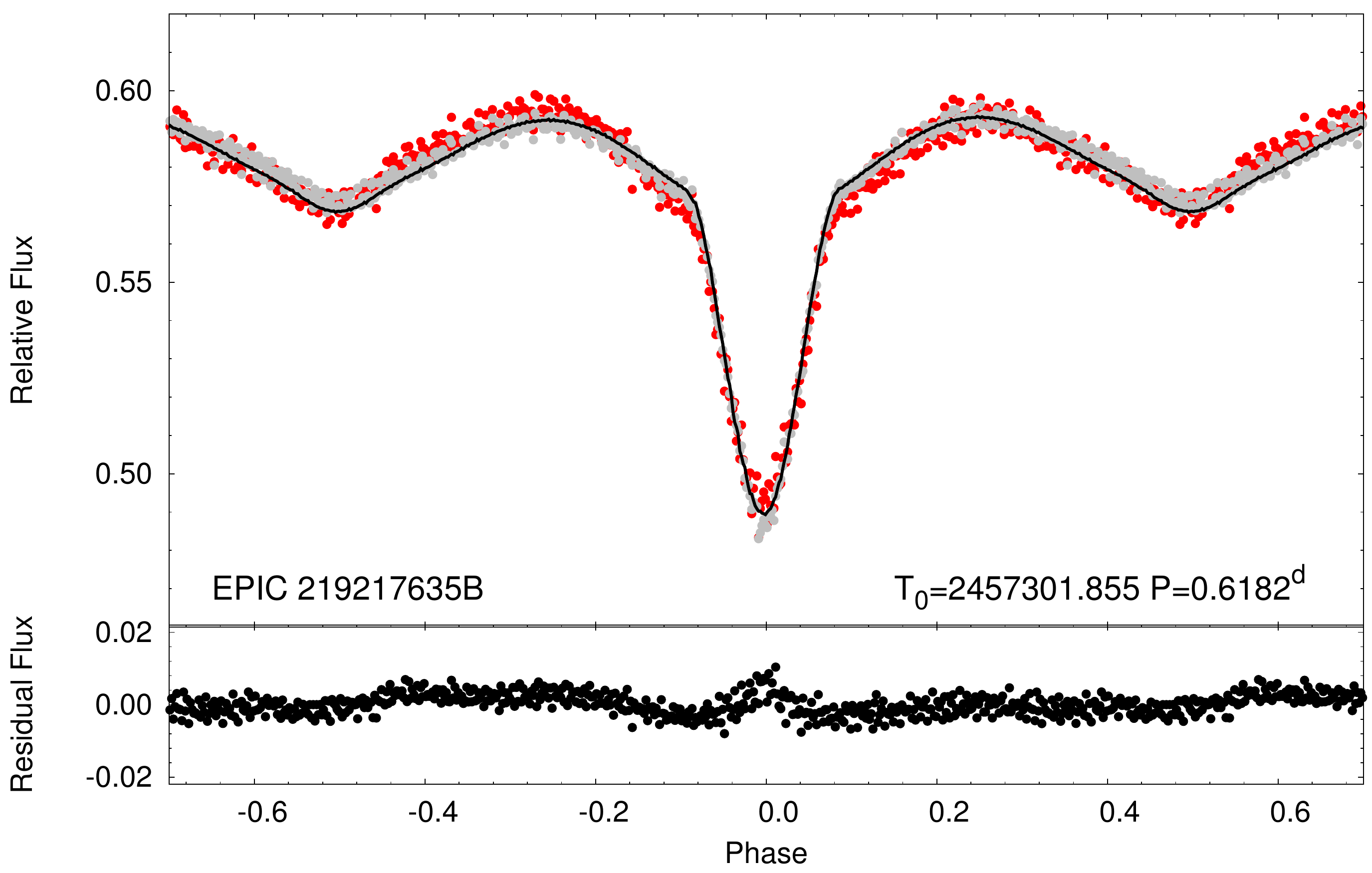}
\caption{The disentangled and folded lightcurves of the 3.595-day `A' binary and the 0.618-day `B' binary (red dots). The black curves represent the disentangled, folded lightcurves, obtained from the simultaneous lightcurve solution (partially shown in Fig.\,\ref{fig:zoomLC}). In the case of binary B the grey dots represent the sum of the disentangled, folded lightcurve and a simple model of the rotational spot modulation (see text for details).  The bottom panel for each binary shows the folded, disentangled residuals of the full K2 data from the model fit.}
\label{fig:folds}
\end{center}
\end{figure}

The disentangled and folded lightcurve of each binary is shown separately in Fig.~\ref{fig:folds}. These plots demonstrate the likely semi-detached nature of the 0.618-day binary and the detached nature of the 3.595-day binary.

We return to a more detailed quantitative analysis of the lightcurves of the two binaries in Section \ref{sec:model}. To start, we simply collect the available photometry on the target-star image in Table \ref{tbl:mags}.  Note that these magnitudes refer to the combined light from all four stars in both binaries.

\begin{table}
\centering
\caption{Properties of the EPIC 219217635 System}
\begin{tabular}{lc}
\hline
\hline
RA (J2000) & 18:59:00.625   \\  
Dec (J2000) &  $-17:15:57.13$  \\  
$K_p$ & 12.72  \\
$B^b$ & 13.86 \\
$g^a$ & 13.42  \\
$V^b$ & 13.13  \\
$R^b$ & 11.74 \\
$r^a$ & 12.72  \\  
$z^a$ & 13.42 \\
$i^b$ & 12.43 \\
J$^c$ & 11.44  \\
H$^c$ & 11.11 \\
K$^c$ & 11.02  \\
W1$^d$ & 10.58 \\
W2$^d$ & 10.61 \\
W3$^d$ & 10.79  \\
W4$^d$ & ... \\
Distance (pc)$^e$ & $870 \pm 100$ \\   
$\mu_\alpha$ (mas ~${\rm yr}^{-1}$)$^f$ & $-1.9 \pm 1.5$  \\ 
$\mu_\delta$ (mas ~${\rm yr}^{-1}$)$^f$ &  $-7.1 \pm 2.4$  \\ 
\hline
\label{tbl:mags}
\end{tabular}

{\bf Notes.} (a) Taken from the SDSS image \citep{Ahn}. (b) From VizieR \url{http://vizier.u-strasbg.fr/}; UCAC4 \citep{UCAC4}. (c) 2MASS catalog \citep{Skrutskie}.  (d) WISE point source catalog \citep{Cutri}. (e) Based on photometric parallax only (see Sect.~\ref{sec:model}).  This utilized an adapted V magnitude of 13.1. (f) From UCAC4 \citep{UCAC4}; \citet{Smart}; \citet{Huber}. 
\end{table}




\section{Adaptive Optics Imaging}
\label{sec:Keck_AO}

We obtained Keck II/NIRC2 (PI: Keith Matthews) observations of the target star EPIC 219217635 on 2017 May 10 UT using the narrow camera ($10'' \times 10''$ field of view) to better characterize this quadruple system. Our observations used the target star as the guide star and dome flat fields and dark frames to calibrate the images and remove artifacts. 

\begin{figure}
\begin{center}
\includegraphics[width=0.45 \textwidth]{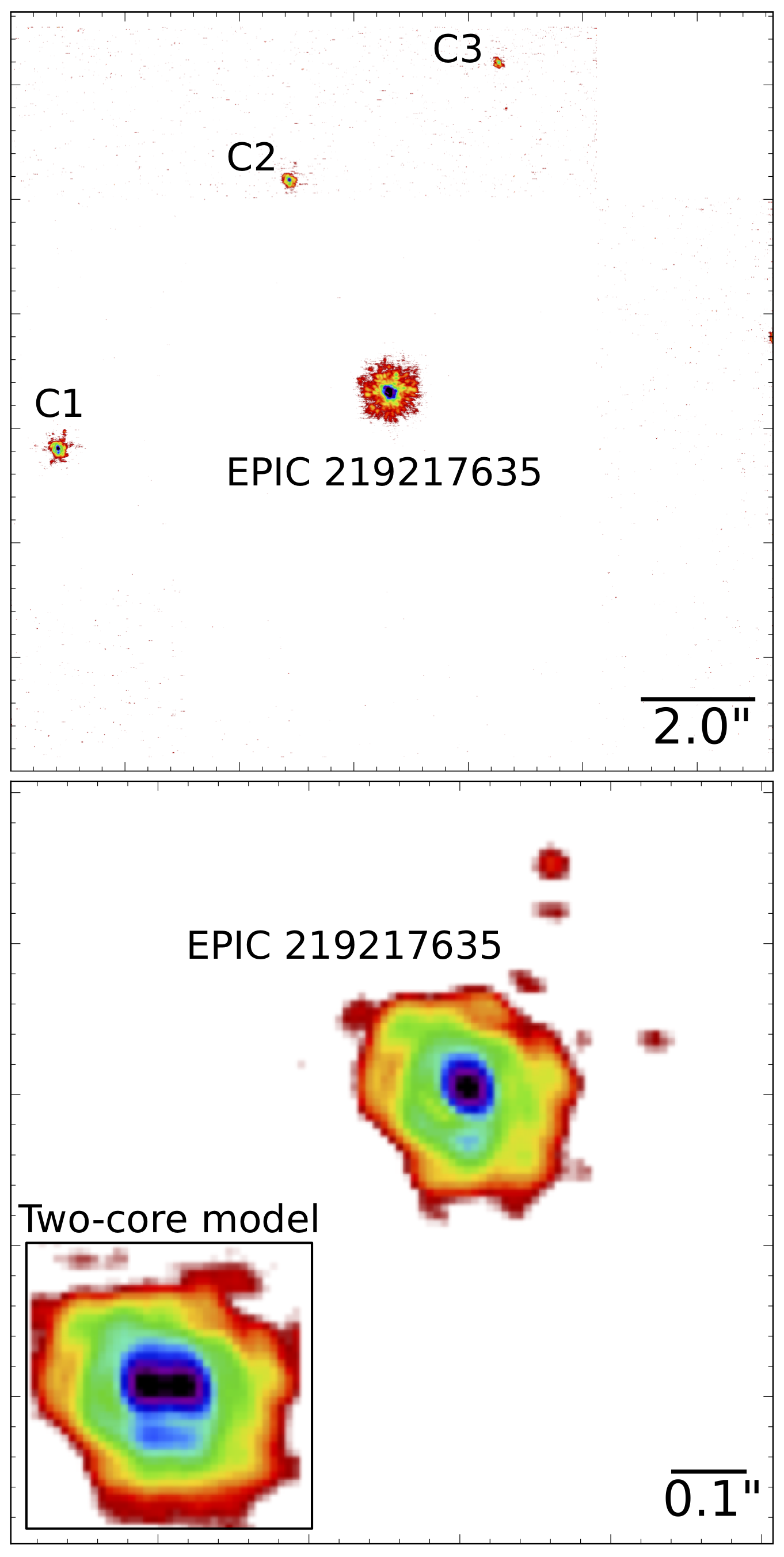}  
\caption{Keck-AO image in $K_s$-band of EPIC 219217635. {\em Top panel}: Full image covering $\sim$$13'' \times 13''$.  Three of the neighboring stars, are labeled C1, C2, and C3 for reference.
{\em Bottom panel}: Zoom-in around the target star EPIC 219217635. The inset shows a simple simulation of what the image would look like if the two binaries were separated by 0.05$''$ (see text for a description of how this was generated). We conclude that the two binaries in this target image are clearly unresolved at the $0.05''$ level.}
\label{fig:Keck_AO}
\end{center}
\end{figure}

We used a 3-point dither pattern to acquire twelve eight-second frames of EPIC 219217635 in the $K_s$ band (central wavelength 2.145 $\mu$m), for a total on-sky integration time of 96 seconds. Figure \ref{fig:Keck_AO} shows a stacked $K_s$ band image of this target 
The top panel shows the full AO image which covers $13'' \times 13''$ on the sky, and includes three of the neighbor stars (labeled C1, C2, and C3), which are likely to be background stars rather than gravitationally bound companions.  The AO photometry for the three nearby stars are given in Table \ref{tbl:Keck_AO}. Due to the large separations of these neighbor stars, C1 only appears in two out of the three dither positions while C2 and C3 appear in only one out of three dither positions. The $K_s$ band astrometry was computed via PSF fitting using a combined Moffat and Gaussian PSF model following the techniques described in \citet{Ngo} and the NIRC2 narrow camera plate scale and distortion solution presented in \citet{Service}.

In the bottom panel of Fig.~\ref{fig:Keck_AO}, we show a zoomed-in image of the target star.  This blown-up image looks distinctly single, and shows no sign of the core even being elongated.  We have carried out simulations of close pairs of comparably bright images, at a range of spacings, and we conclude from this that separations between the two binaries of $\gtrsim 0.05''$ can be conservatively ruled out.  At a source distance of some 870 pc, this sets an upper limit on the projected physical separation of $\sim$50 AU.

A simple demonstration of what the AO image would look like if the two binaries (of nearly equal brightness; see Sect.~\ref{sec:model}) were separated by 0.05$''$ in the horizontal direction is shown in the inset to the bottom panel in Fig.~\ref{fig:Keck_AO}. To generate the inset figure, we simply duplicated the zoomed-in AO image, shifted it by 0.05$''$ in the horizontal direction, and added it to the original image.  One can see that if the two binaries were indeed separated by 0.05$''$, the core of the image would be noticeably elongated.

\begin{table}
\centering
\caption{Stellar Neighbors of EPIC 219217635$^a$}
\begin{tabular}{lcccc}
\hline
Star & Flux Ratio & Separation & Pos.~Angle & $t_{{\mathrm{exp}}^{(b)}}$\\
       &  (Ks band) & (mas) & (deg E of N)  & (s) \\
\hline
C1 & $5.18 \pm 0.11$  & $5873 \pm 2.9$ & $99.53 \pm 0.03$ & 64\\
C2 & $11.49 \pm 0.58$ & $4087 \pm 2.2$ & $25.08 \pm 0.03$ & 32\\
C3 & $29.75 \pm 0.74$  & $6036 \pm 3.0$ & $341.39 \pm 0.03$ & 32\\
\hline
\label{tbl:Keck_AO}
\end{tabular}

{\bf Notes.} (a) Results obtained from the Keck AO image. (b) Total exposure time on each neighbor star. While the target star was present for the full 96 seconds of integration, the neighbor stars only appeared in-frame for a subset of the dither positions.
\end{table}

\begin{table*}
\caption{Mid-times of primary eclipses of EPIC 219217635A}
 \label{Tab:EPIC_219217635A_ToM}
\begin{tabular}{@{}lrllrllrl}
\hline
BJD & Cycle  & std. dev. & BJD & Cycle  & std. dev. & BJD & Cycle  & std. dev. \\ 
$-2\,400\,000$ & no. &   \multicolumn{1}{c}{$(d)$} & $-2\,400\,000$ & no. &   \multicolumn{1}{c}{$(d)$} & $-2\,400\,000$ & no. &   \multicolumn{1}{c}{$(d)$} \\ 
\hline
57302.37640 &	-1.0 & 0.00057 & 57334.73057 &    8.0 & 0.00012 & 57367.08298 &   17.0 & 0.00129 \\ 
57305.97293 &	 0.0 & 0.00014 & 57338.32397 &    9.0 & 0.00022 & 57370.67604 &   18.0 & 0.00019 \\ 
57309.56653 &	 1.0 & 0.00018 & 57341.92041 &   10.0 & 0.00016 & 57374.27121 &   19.0 & 0.00019 \\ 
57313.16229 &	 2.0 & 0.00019 & 57345.51450 &   11.0 & 0.00017 & 57377.86827 &   20.0 & 0.00173 \\ 
57316.75708 &	 3.0 & 0.00020 & 57349.10967 &   12.0 & 0.00017 & 57381.46098 &   21.0 & 0.00012 \\\cline{7-9} 
57320.35134 &	 4.0 & 0.00019 & 57352.70271 &   13.0 & 0.00046 & 57891.91419 &  163.0 & 0.00009 \\ 
57323.94668 &	 5.0 & 0.00022 & 57356.29934 &   14.0 & 0.00031 & 57924.26923 &  172.0 & 0.00010 \\ 
57327.54058 &	 6.0 & 0.00018 & 57359.89390 &   15.0 & 0.00011 & 57942.24434 &  177.0 & 0.00020 \\
57331.13552 &	 7.0 & 0.00039 & 57363.48917 &   16.0 & 0.00022 & 57988.97864 &  190.0 & 0.00025 \\
\hline
\end{tabular}

{\bf Notes.} Most of the eclipses (cycle nos. $-1-21$) were observed by {\em Kepler} spacecraft. Last four eclipses (under the horizontal line) were observed at HAO (no. 163) and PEST (nos. $172-190$)  observatories.
\end{table*}

\begin{table*}
\caption{Mid-times of primary eclipses of EPIC 219217635B}
 \label{Tab:EPIC_219217635B_ToM}
\begin{tabular}{@{}lrllrllrl}
\hline
BJD & Cycle  & std. dev. & BJD & Cycle  & std. dev. & BJD & Cycle  & std. dev. \\ 
$-2\,400\,000$ & no. &   \multicolumn{1}{c}{$(d)$} & $-2\,400\,000$ & no. &   \multicolumn{1}{c}{$(d)$} & $-2\,400\,000$ & no. &   \multicolumn{1}{c}{$(d)$} \\ 
\hline
57301.85339 &	 0.0 & 0.00055 & 57329.67001 &   45.0 & 0.00027 & 57357.49055 &	90.0& 0.00043 \\
57303.08879 &	 2.0 & 0.00020 & 57330.28886 &   46.0 & 0.00107 & 57358.72688 & 92.0& 0.00047 \\ 
57303.70704 &	 3.0 & 0.00045 & 57330.90695 &   47.0 & 0.00022 & 57359.34632 & 93.0& 0.00036 \\ 
57304.32550 &	 4.0 & 0.00027 & 57331.52407 &   48.0 & 0.00039 & 57360.58266 & 95.0& 0.00050 \\ 
57304.94365 &	 5.0 & 0.00027 & 57332.14301 &   49.0 & 0.00007 & 57361.20038 & 96.0& 0.00012 \\ 
57305.56253 &	 6.0 & 0.00152 & 57332.76132 &   50.0 & 0.00025 & 57362.43609 & 98.0& 0.00067 \\ 
57306.18046 &	 7.0 & 0.00003 & 57333.37931 &   51.0 & 0.00083 & 57363.05545 & 99.0& 0.00047 \\ 
57306.79734 &	 8.0 & 0.00071 & 57333.99694 &   52.0 & 0.00061 & 57363.67455 &100.0& 0.00061 \\ 
57307.41647 &	 9.0 & 0.00048 & 57334.61525 &   53.0 & 0.00014 & 57364.29122 &101.0& 0.00017 \\ 
57308.03398 &	10.0 & 0.00091 & 57335.23419 &   54.0 & 0.00078 & 57364.91057 &102.0& 0.00114 \\ 
57308.65247 &	11.0 & 0.00076 & 57335.85254 &   55.0 & 0.00026 & 57365.52781 &103.0& 0.00080 \\ 
57309.27014 &	12.0 & 0.00022 & 57337.08857 &   57.0 & 0.00031 & 57366.14836 &104.0& 0.00045 \\ 
57309.88869 &	13.0 & 0.00108 & 57337.70643 &   58.0 & 0.00032 & 57366.76422 &105.0& 0.00031 \\ 
57310.50691 &	14.0 & 0.00012 & 57338.94273 &   60.0 & 0.00006 & 57367.38405 &106.0& 0.00077 \\ 
57311.12496 &	15.0 & 0.00055 & 57339.56161 &   61.0 & 0.00011 & 57368.00177 &107.0& 0.00084 \\ 
57311.74309 &	16.0 & 0.00007 & 57340.79859 &   63.0 & 0.00075 & 57368.62095 &108.0& 0.00047 \\ 
57312.36134 &	17.0 & 0.00009 & 57342.03502 &   65.0 & 0.00036 & 57369.23842 &109.0& 0.00038 \\ 
57312.97808 &	18.0 & 0.00032 & 57342.65434 &   66.0 & 0.00089 & 57369.85815 &110.0& 0.00200 \\ 
57313.59722 &	19.0 & 0.00026 & 57343.27132 &   67.0 & 0.00023 & 57370.47323 &111.0& 0.00026 \\ 
57314.21555 &	20.0 & 0.00036 & 57343.89018 &   68.0 & 0.00025 & 57371.09366 &112.0& 0.00041 \\ 
57314.83386 &	21.0 & 0.00009 & 57344.50773 &   69.0 & 0.00024 & 57371.71162 &113.0& 0.00039 \\ 
57315.45107 &	22.0 & 0.00316 & 57345.12682 &   70.0 & 0.00090 & 57372.33037 &114.0& 0.00067 \\ 
57316.07058 &	23.0 & 0.00034 & 57345.74510 &   71.0 & 0.00041 & 57372.94706 &115.0& 0.00044 \\ 
57317.30683 &	25.0 & 0.00010 & 57346.36355 &   72.0 & 0.00078 & 57373.56618 &116.0& 0.00028 \\ 
57317.92490 &	26.0 & 0.00070 & 57346.98041 &   73.0 & 0.00011 & 57374.80215 &118.0& 0.00045 \\ 
57319.16173 &	28.0 & 0.00057 & 57347.59947 &   74.0 & 0.00042 & 57375.41912 &119.0& 0.00070 \\ 
57321.01640 &	31.0 & 0.00040 & 57348.21883 &   75.0 & 0.00107 & 57376.65586 &121.0& 0.00086 \\ 
57321.63400 &	32.0 & 0.00006 & 57348.83568 &   76.0 & 0.00023 & 57377.27327 &122.0& 0.00047 \\ 
57322.87125 &	34.0 & 0.00014 & 57349.45359 &   77.0 & 0.00013 & 57378.51223 &124.0& 0.00009 \\ 
57323.48879 &	35.0 & 0.00046 & 57350.07298 &   78.0 & 0.00040 & 57379.12853 &125.0& 0.00041 \\ 
57324.10676 &	36.0 & 0.00063 & 57350.69137 &   79.0 & 0.00037 & 57380.36682 &127.0& 0.00074 \\ 
57324.72549 &	37.0 & 0.00035 & 57351.30971 &   80.0 & 0.00014 & 57380.98310 &128.0& 0.00066 \\ 
57325.34337 &	38.0 & 0.00029 & 57351.92734 &   81.0 & 0.00137 & 57381.60127 &129.0& 0.00047 \\ 
57325.96183 &	39.0 & 0.00059 & 57352.54397 &   82.0 & 0.00022 & 57382.21900 &130.0& 0.00029 \\\cline{7-9}
57326.57965 &	40.0 & 0.00053 & 57353.78200 &   84.0 & 0.00037 & 57910.16115 &984.0& 0.00013 \\ 
57327.19787 &	41.0 & 0.00034 & 57355.01923 &   86.0 & 0.00029 & 57923.14305 &1005.0&0.00015 \\ 
57327.81569 &	42.0 & 0.00039 & 57355.63804 &   87.0 & 0.00178 & 57924.38240 &1007.0&0.00023 \\ 
57328.43373 &	43.0 & 0.00034 & 57356.87288 &   89.0 & 0.00035 & 57929.32426 &1015.0&0.00013 \\ 
57329.05172 &   44.0 & 0.00054 &             &	      &	        &             &     &         \\ 
\hline
\end{tabular}

{\bf Notes.} Most of the eclipses (cycle nos. $0-130$) were observed by {\em Kepler} spacecraft. Last four eclipses (under the horizontal line) were observed at the PEST observatory.
\end{table*}


\section{Ground-based photometry}
\label{sec:photometry}

\subsection{HAO Observations}

The Hereford Arizona Observatory (HAO) consists of a 0.34-m Meade brand Schmidt-Cassegrain telescope (`SCT') on a fork mount, inside an ExploraDome. All hardware is controlled via buried cables from a nearby residence. Maxim DL 5.2 software is used to control the telescope, dome, focuser, filter wheel and SBIG ST-10XME CCD camera. The unbinned image scale was 0.52$''$/pixel. All observations were made using a V-band filter, with exposure times of 60 seconds. Images were calibrated using master bias, dark and flat images. Ten reference stars and 7 calibration stars were employed for converting instrument magnitude to V-magnitude. 

\subsection{PEST Observations}

PEST is a home observatory with a 12-inch Meade LX200 SCT f/10 telescope with a SBIG ST-8XME CCD camera. The observatory is owned and operated by Thiam-Guan (TG) Tan. PEST is equipped with a BVRI filter wheel, a focal reducer yielding f/5, and an Optec TCF-Si focuser controlled by the observatory computer. PEST has a $31' \times 21'$ field of view and a $1.2''$ per pixel scale.  PEST is located in a suburb of the city of Perth, Western Australia.  PEST observed EPIC 219217635 on 7 nights between June 5, 2017 and August 23, 2017 in the V band with 120-s integration times.

In all, the HAO and PEST observations led to measurements of four precise primary eclipse times for the 3.595-day A binary and an equal number of primary eclipses for the 0.618-day B binary (see the last columns of Tables\,\ref{Tab:EPIC_219217635A_ToM} and \ref{Tab:EPIC_219217635B_ToM}). Additionally, on the night of June 12, 2017 an event involving an overlapping primary eclipse of binary A and a secondary eclipse of binary B was also observed at PEST Observatory.  However, due to the composite nature of this eclipse we were not able to determine the mid-eclipse times with satisfactory accuracies and, therefore, we did not tabulate this event.

\section{Period study}   
\label{sec:ETV}

In order to look for and analyse the possible eclipse timing variations (`ETVs') in the two binaries, we determined the times of each eclipse minimum using the K2 data with the blended binaries in the following manner.  First we formed a folded, binned lightcurve with the period of the 0.618-day binary B in such a way that the narrow region around the primary and secondary eclipses of the 3.595-day binary A were omitted. Then, the profile of the primary eclipse of this folded lightcurve (lower panel of Fig.\,\ref{fig:folds}) was used as a template for calculating the times of the primary eclipses of binary B in the K2 dataset. (We decided not to utilize the secondary eclipses, due to the fact that they are rather shallow.)  

In order to obtain the times of the primary eclipses of binary A, we removed the folded, binned, averaged binary B lightcurve from the K2 dataset with the use of a three-point local Lagrange interpolation. Then, this disentangled lightcurve (upper panel of Fig.\,\ref{fig:folds}) was used both for forming the folded, binned, averaged lightcurve of binary A and, also for determining the times of the primary eclipses of binary A. (Here, for the same reasons as mentioned above, we utilized only the times of the primary eclipses.) 

\begin{figure*}
\begin{center}
\includegraphics[width=0.70 \textwidth]{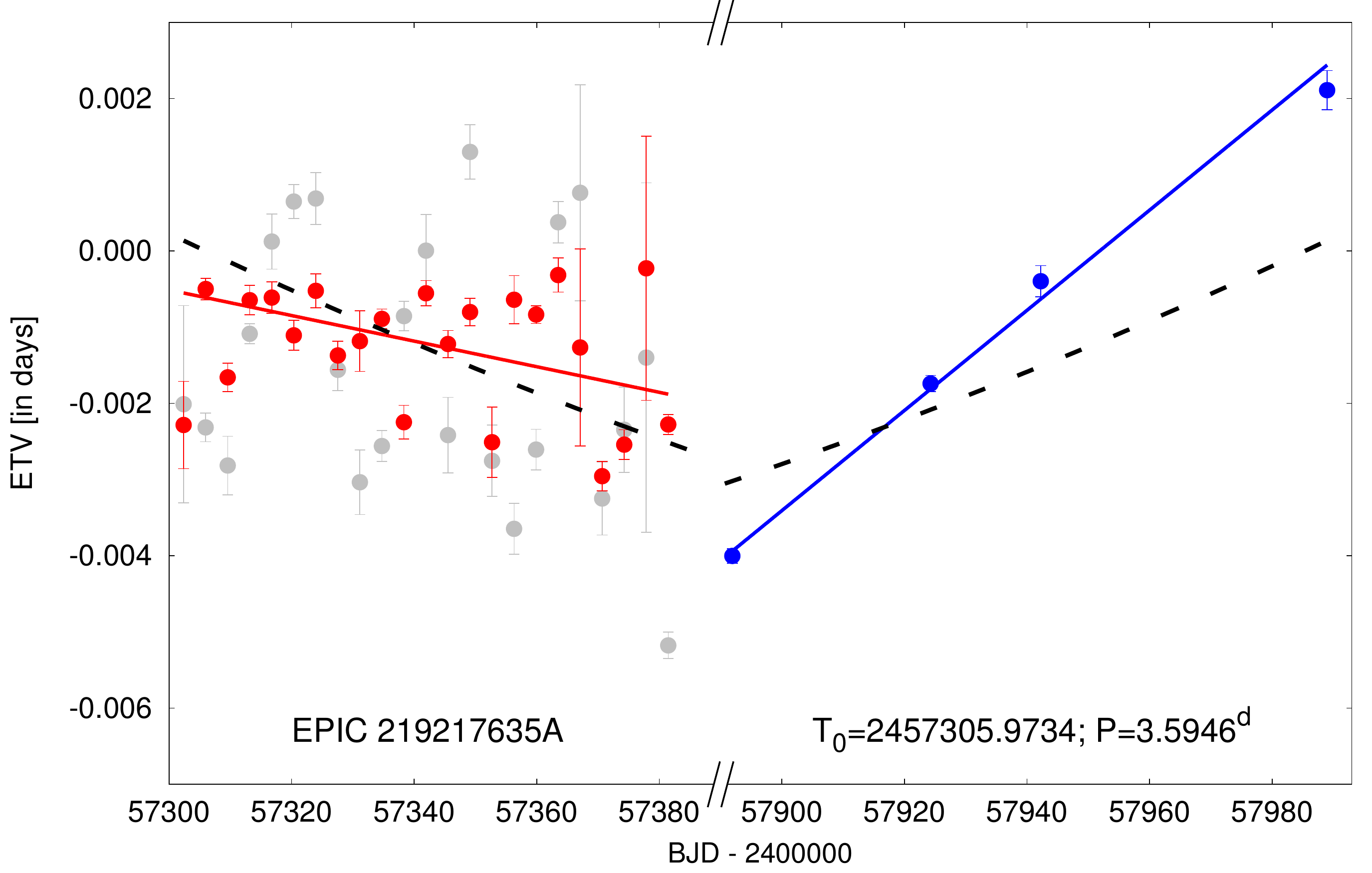}
\caption{Eclipse Timing Variations of binary A. Grey and red circles represent eclipse times determined from the K2 observations before and after the removal of non-binary lightcurve variations, respectively. Blue data points represent the ground-based timing measurements. Red and blue lines are linear fits to the red and blue ETV points, respectively, which would illustrate two constant-period segments with a period difference of 26\,sec (note in particular the broken time axis). Black dashed lines illustrate the two sections of a parabola that result from a quadratic fit to the red and blue ETV points together, i.e. modeling a constant rate of increase in the orbital period.}
\label{fig:ETV-A}
\end{center}
\end{figure*}

\begin{figure*}
\begin{center}
\includegraphics[width=0.70 \textwidth]{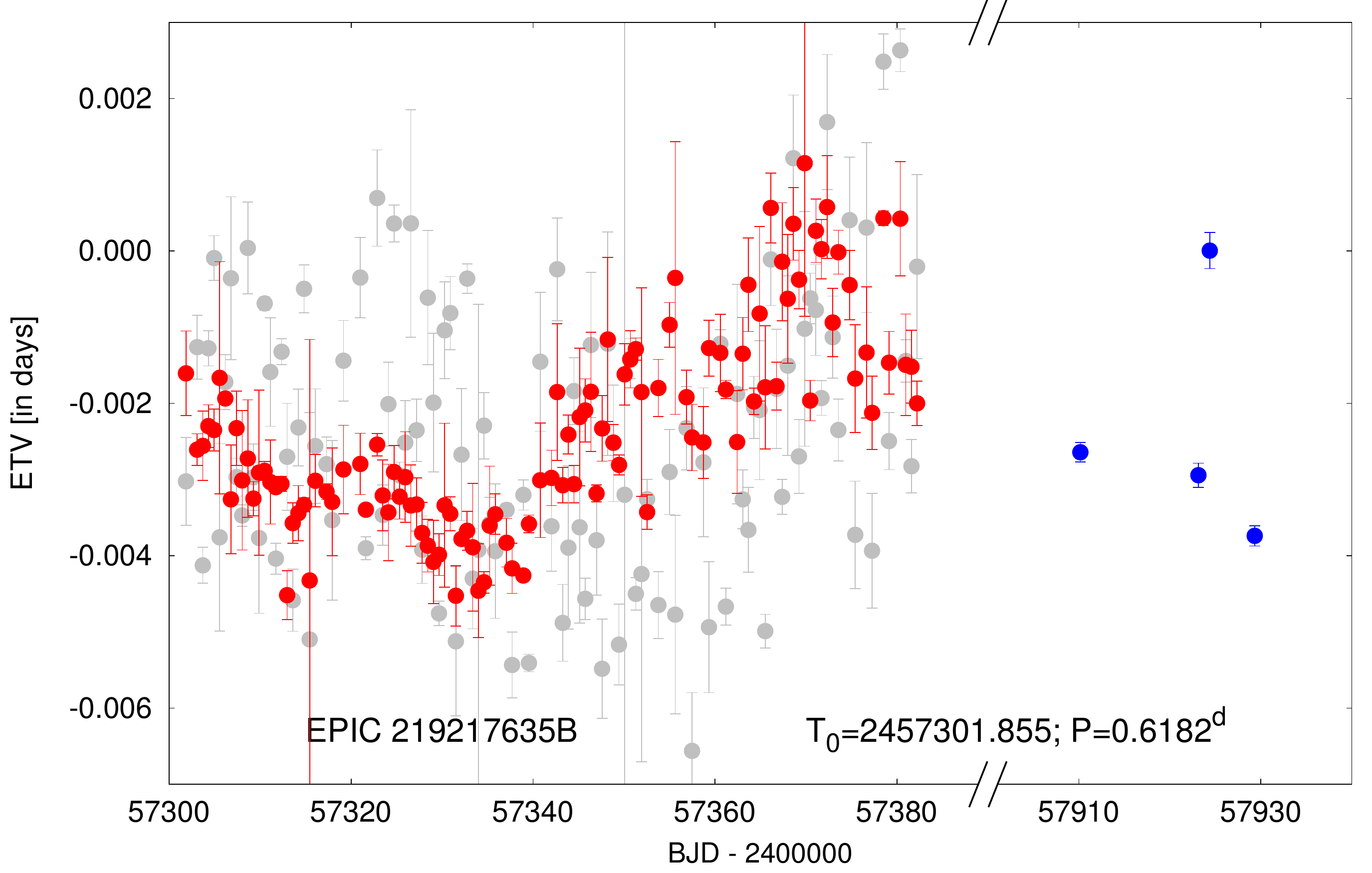}
\caption{Eclipse Timing Variations of binary B. Grey and red circles represent eclipse times determined from the K2 observations before and after the removal of lightcurve variations, respectively, that are not inherent to the binary lightcurve. Blue data points represent the ground-based timing measurements.  Note the broken time axis between the two sets of observations.}
\label{fig:ETV-B}
\end{center}
\end{figure*}

In such a way we obtained the first-iteration K2 ETV curves for both binaries. Later, however, during our analysis, we realized that besides the classical binary lightcurve variations, the lightcurve also exhibits some additional periodic variations (see Sect.\,\ref{sec:model}). Thus, after the separation and removal of these extra periodic signals from the K2 lightcurve, we repeated the process described above, and we were able to refine the ETV curves (see, Figs.\,\ref{fig:ETV-A} and \ref{fig:ETV-B}, and also Tables\,\ref{Tab:EPIC_219217635A_ToM} and \ref{Tab:EPIC_219217635B_ToM}). 

Furthermore, we have carried out ground-based photometric follow up observations with two telescopes on eight nights between May and August 2017 (see Sect.~\ref{sec:photometry}). We were thereby able to determine 8 additional primary eclipse times (4 for both binaries; given at the end of Tables\,\ref{Tab:EPIC_219217635A_ToM} and \ref{Tab:EPIC_219217635B_ToM}), which made it possible to extend significantly the observing window and to check for longer timescale trends in the period variations of the two binaries. In order to determine the ground-based eclipse times, we first converted these observations  to the flux regime and then used the same K2 template eclipse profiles as before. Furthermore, in the case of the binary A eclipses we removed the ellipsoidal light variations of binary B via the use of the folded, disentangled binary B K2 lightcurve after phasing it according to its expected phase at the epoch of the ground-based observations.

Regarding the eclipse timings of binary A (Fig.\,\ref{fig:ETV-A}) no definitive short-term ETVs can be seen during the 80 days of the K2 observations. The constant binary period is found to be $P_\mathrm{A-K2}=3\fd59469\pm0\fd00002$. On the other hand, the four ground-based eclipse times (which span a similar time interval) do not phase up to the K2 data. Fitting a constant period to the four ground-based data points yields $P_\mathrm{A-2017}=3\fd59499\pm0\fd00001$ which differs by $\sim$$25.6$\,sec from the K2 period (at the 12-$\sigma$ level). We also fit the joint K2 and 2017 ground-based data using a quadratic ephemeris (see black, dashed segments of the corresponding parabola in Fig.\,\ref{fig:ETV-A}).  A parabolic ETV represents a linear period variation during the 1.9-yr span of both sets of observations. As one can see, the parabolic fit is quite poor. The resultant period variation rate is found to be $\Delta P=1.4\pm0.3\times10^{-6}$\,day/cycle or, $\dot P/P=4.0\pm0.8\times10^{-5}\,\mathrm{yr}^{-1}$. Assuming that the source of this period variation were Keplerian orbital motion of the binary around the center of mass of the quadruple system, one can convert this quantity into a variation in the systemic radial velocity of binary A, as $\dot\gamma_A \approx c\Delta P/P^2$, which results in $\dot\gamma_A  \simeq 0.038\pm0.008\,\mathrm{cm\,s}^{-2}$. As we find later, this value is an order of magnitude higher than we find directly from our radial velocity study (see Sect.~\ref{sec:NOT-FIES}).

We turn now to the ETV curve for binary B (see Fig.~\ref{fig:ETV-B}).  In this case the K2 data, after the removal of the non-binary lightcurve variations, clearly reveal short-term, non-linear behavior in the timing data. On the other hand, however, this non-linear trend, which would correspond to an increasing orbital period, obviously did not continue all the way to the time of the ground-based observations. These latter measurements are in conformity with a constant average period since the beginning of the K2 observations.

Speculating on the origin of these period variations, we can only state with certainty that none of them could arise from the orbit of the two binaries around each other. First, there is the evident contradiction between the period variations found in binary A and the directly measured value of $\dot \gamma_A$ found for binary A (see Sect.~\ref{sec:NOT-FIES}).  Second, there is also the fact that, according to our combined RV and lightcurve solution (see Sect.\,\ref{sec:model}), the total mass of each of the two binaries is similar and, therefore, the ETVs arising from the orbits of the two binaries forming the quadruple system should be similar in amplitude and opposite in phase.\footnote{Strictly speaking this is only true for the light-travel-time effect. The dynamical contribution to the ETVs would differ due to the different periods of the two inner binaries.} In the case of binary B, the spotted nature of at least one of the stars might offer a plausible explanation for the observed short-term ETVs, as similar behaviour has been reported for several spotted {\em Kepler} binaries \citep[see, e.g.][]{tranetal13,balajietal15}.

In the case of binary A, an interpretation of the observed ETV behaviour will require further observations.

\section{NOT-FIES radial velocity study}
\label{sec:NOT-FIES}
 
We obtained 20 spectra of EPIC\,219217635 employing the Nordic Optical
Telescope and its FIES spectrograph \citep{frandsen,telting} in high resolution mode
($R\sim67\,000$). The spectra have been taken between May 18 2016 and
July 05 2017 with exposure times ranging between 20 and 35\,min.
Each science exposure was accompanied by one ThAr exposure immediately
prior for wavelength calibration.

The data reduction was carried out using
FIESTool\footnote{\url{http://www.not.iac.es/instruments/fies/fiestool/}}.
In the following we used the wavelength calibrated extracted, but not
order-merged spectra. Cosmic rays have been identified and removed, the
blaze function of the spectrograph was accounted for using flat-field
exposures, and the spectra have been normalized. For the purpose of
obtaining radial velocities (RV) we focus on the spectral region between
$4500$~\AA{} and $6700$~\AA{}. At shorter wavelengths the typical
signal-to-noise ratio per spectral bin is below 3 for the combined spectrum 
of the two binaries. At longer wavelengths few stellar lines are present. We 
created cross-correlation functions (`CCFs') for each spectral order of each observation 
using a template obtained from the PHOENIX library \citep{Husser}. Specifically we
used the PHOENIX model with $T_{\rm eff} = 6500$~K, $\log \, g = 4.0$ and solar
metallicity. We checked if using different templates with somewhat
different parameters changes the RV we derive (see below), which is
not the case.

Next we fitted two Gaussians to the CCF of each observation obtained by
simple summation of all CCFs from the different orders. One Gaussian
has a small $\sigma$ of $11$~km\,s$^{-1}$ representing the primary from
binary A. The second Gaussian with $\sigma = 120$~km\,s$^{-1}$,
represents the primary from  binary B. The positions of these Gaussians
are interpreted as RVs of the two primary components. We estimate the
uncertainties in these RVs using the following approach. The CCFs from
the different spectral orders are grouped into four different wavelength
regions. RVs for each of the four different orders are obtained in the
same way as for the CCFs from the complete spectral region and the
standard deviation about the mean is used as the RV uncertainty. 

\begin{figure*} 
\begin{center}
\includegraphics[width=0.69 \textwidth]{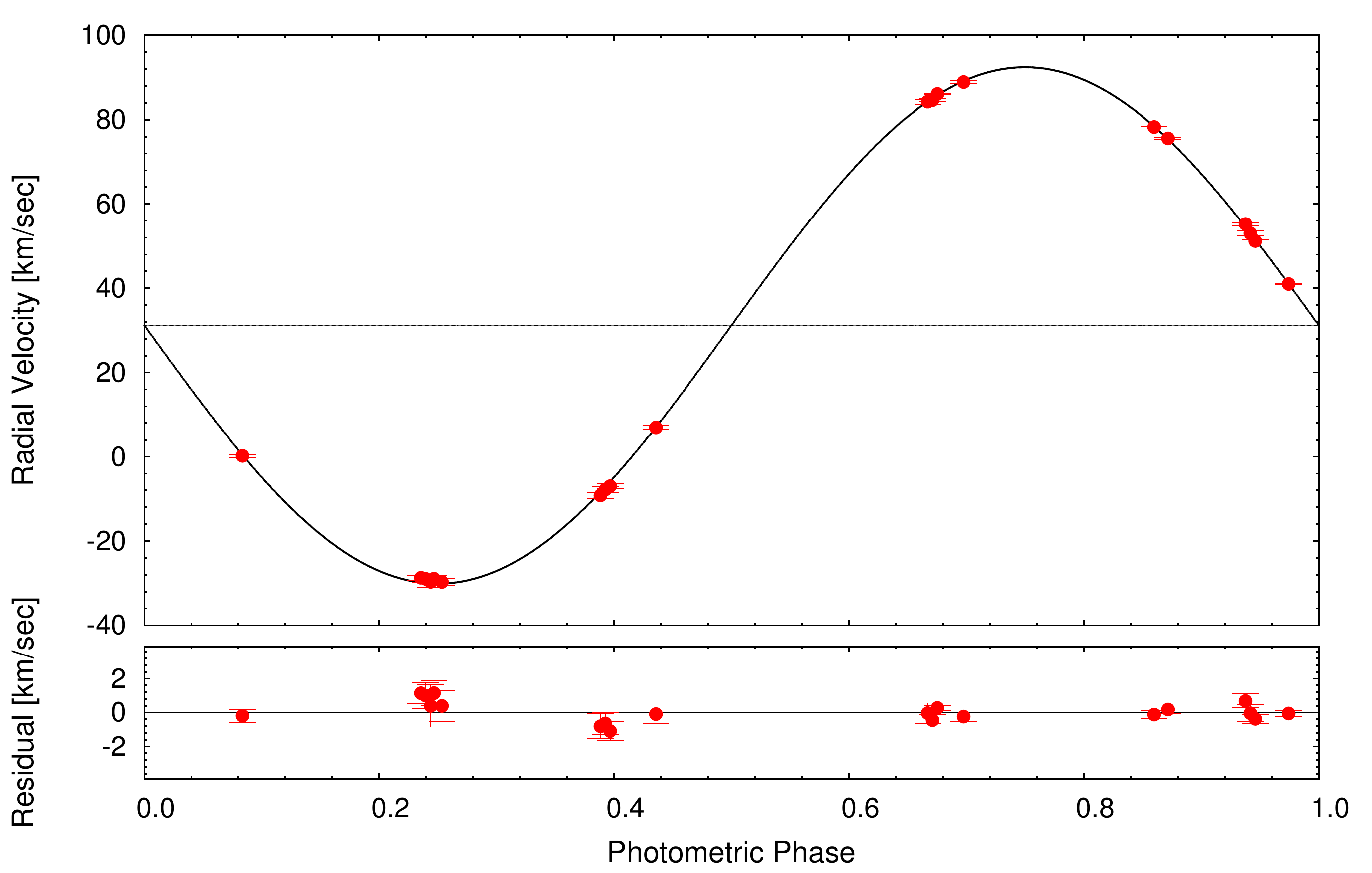} 
\includegraphics[width=0.69 \textwidth]{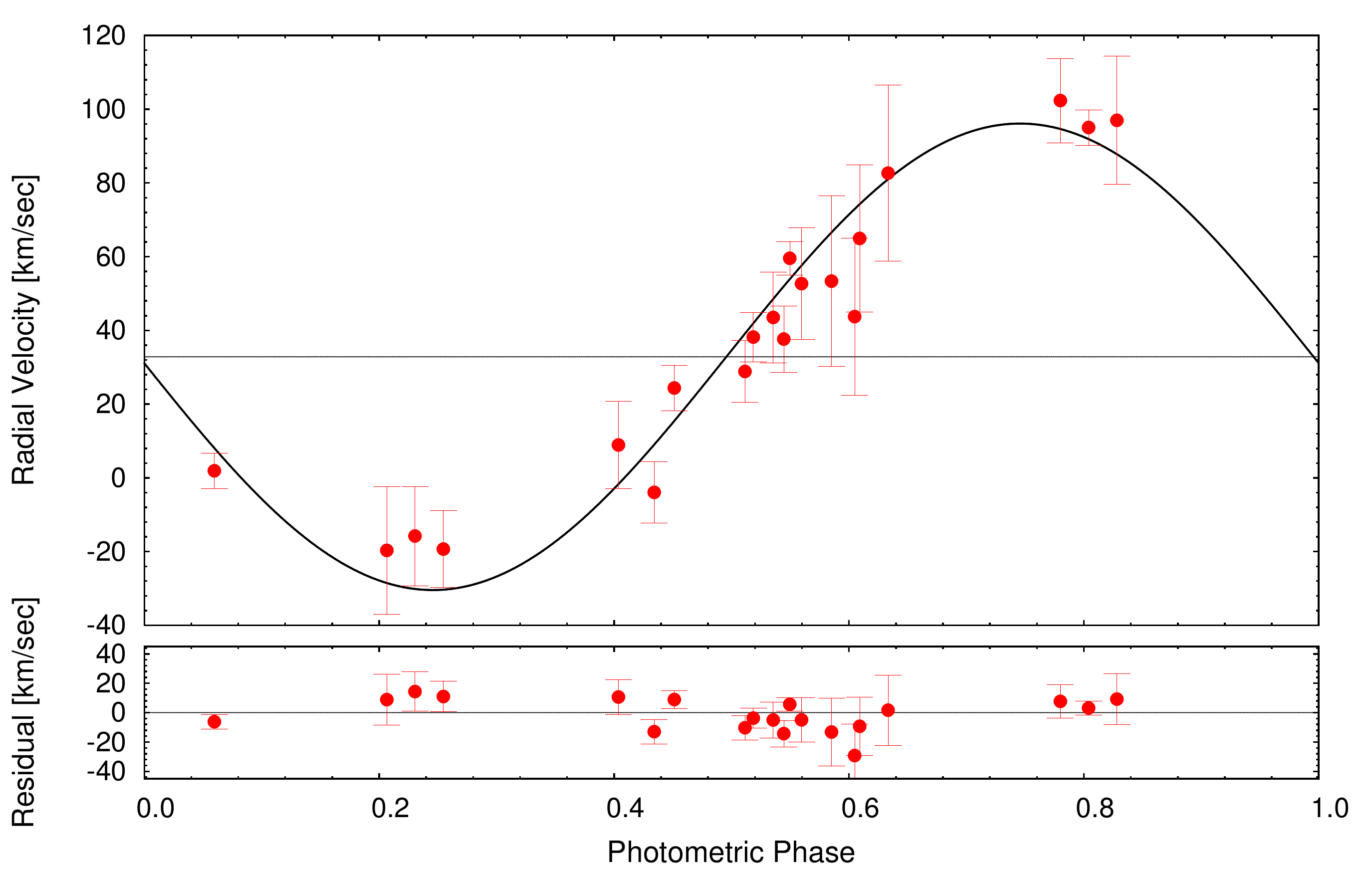} 
\caption{Folded radial velocity curves for the 3.595-day `A' binary (top panel) and the 0.618-day `B' binary (bottom panel). The black curves are the best-fitting circular-orbit models. For a better visualisation all the individual observed and model data points are corrected for the non-zero $\dot\gamma$ values, i.e. we plot the $v_\mathrm{corr}=v_\mathrm{obs/mod}-\dot\gamma\cdot(t_\mathrm{obs/mod}-t_0)$ data.}
\label{fig:RVs}
\end{center}
\end{figure*}

The radial velocity plots obtained with the NOT-FIES spectrometer are shown in Fig.~\ref{fig:RVs}, and the individual RV measurements are listed in Table \ref{tbl:RVs}.  The RV curve for the primary star in binary A (top panel) has very well determined parameter values with a typical uncertainty per RV point of $\sim$0.5 km s$^{-1}$.  The orbital amplitude, $K_A$, is $61.28 \pm 0.15$ km s$^{-1}$, while the system velocity is $\gamma_A = 30.91 \pm 0.13$ km s$^{-1}$.  For the primary component in binary B (bottom panel), the typical uncertainties per RV point are $\sim$12 km s$^{-1}$.  The corresponding elements are: $K_B = 64.1 \pm 3.6$ km s$^{-1}$ and $\gamma_B = 32.3 \pm 2.2$ km s$^{-1}$. These were all for assumed circular orbits, but we fit for, and set constraints on, eccentric orbits as well.

\begin{figure}
\begin{center}
\includegraphics[width=0.49 \textwidth]{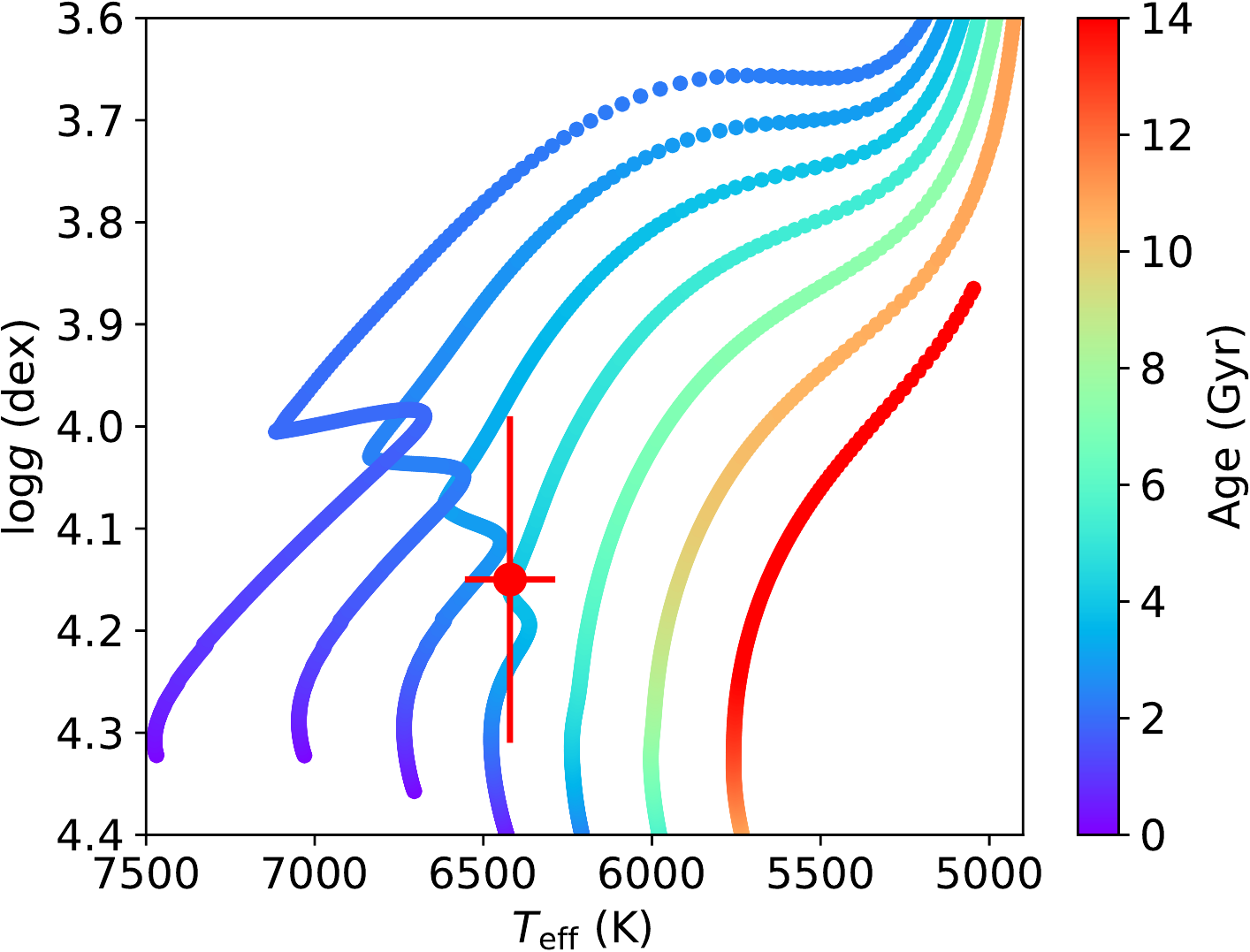}
\caption{The spectroscopically determined location (with uncertainties) of star A1 in the $\log \, g-T_{\rm eff}$ plane.  The colored curves are evolution tracks for stars of mass 0.9 to 1.5\,M$_\odot$ (increasing from left to right) in steps of 0.1\,M$_\odot$.  Tracks are color coded according to the isochrones of stellar evolution time.  See text for details.}
\label{fig:tracks}
\end{center}
\end{figure}  

\begin{table*}
\centering
\caption{Radial Velocity Study$^a$}
\begin{tabular}{lcc}
\hline
\hline
 & Binary A & Binary B  \\
\hline
RV Measurements: & & \\
\hline
BJD-2400000 & km s$^{-1}$ & km s$^{-1}$ \\
\hline
57526.6406	& $-9.40 \pm 0.73$ & $+55.8 \pm 23$ \\			
57526.6554	& $-8.01 \pm 0.64$  & $+67.3 \pm 20$ \\			
57526.6703	& $-7.20 \pm 0.55$ & $+85.1	 \pm 24$ \\			
57527.6428	& $+84.09 \pm 0.59$ & $-17.3 \pm 17$ \\			
57527.6576	& $+84.45  \pm 0.35$ &$-13.4 \pm 13$ \\			
57527.6725	& $+85.91 \pm 0.16$ & $-17.0 \pm 10$ \\			
57528.6155	& $+55.04 \pm 0.41$ & $+104.7 \pm 11$ \\			
57528.6303	& $+52.84 \pm 0.51$ & $+97.4 \pm 5$ \\			
57528.6452	& $+51.01 \pm 0.26$ & $+99.3 \pm 17$ \\			
57529.6858	& $-28.91  \pm 0.60$ & $+31.2 \pm 8$ \\			
57529.7006	& $-29.18  \pm 0.77$ & $+45.8 \pm 12$ \\			
57529.7155	& $-29.87  \pm 1.24$ & $+55.0 \pm 15$ \\		
57666.3547	& $-29.62  \pm 0.91$ & $+42.7 \pm 21$ \\			
57669.3399	& $+0.31   \pm 0.37$ & $-5.0 \pm 8$ \\			
57682.3316	& $+89.05  \pm 0.27$ & $+23.0 \pm 6$ \\		
57683.3258	& $+41.12 \pm 0.19$ & $+0.53 \pm 5$ \\			
57864.7275	& $+7.46   \pm 0.53$ & $+32.5 \pm 7$ \\			
57916.5810	& $+78.88  \pm 0.22$ & $+1.94 \pm 12$ \\			
57934.5976	& $+76.21 \pm 0.26$ & $+52.1 \pm 5$ \\		
57939.5396 & $-28.29  \pm 0.75$ & $+30.1 \pm 9$ \\		
\hline
Orbit Fits: & \\
\hline
$T_0$ [BJD]$^b$ & $2457625.9007 \pm 0.0012$ & $2457625.801 \pm 0.005$ \\
P [days] & 3.59486(4)  &  0.61815(2) \\ 
K [km s$^{-1}$] & $61.28 \pm 0.15$ & $64.1 \pm 3.6$  \\
$\gamma$ [km s$^{-1}$]   &  $+30.91 \pm 0.13$    &  $+32.3 \pm 2.2$  \\
$e$ & $\la 0.01$ & ... \\
$\dot{\gamma}$ [cm s$^{-2}$]$^c$   & $0.0024 \pm 0.0007$   & $-0.020 \pm 0.014$ \\
\hline
Spectroscopic Parameters$^d$: & & \\
\hline
$T_{\rm eff}$ [K] & $6421 \pm 134$ & ... \\
$\log \, g$ [cgs] & $4.15 \pm 0.16$$^e$ & ... \\
Fe/H [dex] & $-0.03 \pm 0.07$ & ...  \\  
$v \, \sin i$ [km s$^{-1}$] & $16.7 \pm 1$ & ... \\
$M_{A1}$ [$M_\odot $] & $1.23^{+0.10}_{-0.08}$ & ... \\
$R_{\rm A1}$ [$R_\odot$] & $1.39^{+0.31}_{-0.17}$ & ... \\
age [Gyr] & $2.4 \pm 1$ & ... \\
\hline
\end{tabular}
\label{tbl:RVs}

{\bf Notes.} (a) Carried out with the NOT-FIES spectrometer. (b) Time of the primary eclipse and reference time for $P$ and $K$. (c) Parameter fitted to the unfolded RV data set. (d) Parameters refer to the primary star which contributes $\gtrsim 90\%$ of the light from the A binary.  (e) Derived from the summed spectra; see Sect.~\ref{sec:NOT-FIES}

\end{table*}

We also used the NOT-FIES spectral data to determine some of the properties of the primary star in binary A.  The results are given in Table \ref{tbl:RVs} and shown in Fig.~\ref{fig:tracks}.  After obtaining RVs for the A and B binaries we use the tomography algorithm developed by \citet{Bagnuolo} to separate the spectra. We stack the separated spectra to obtain coadded, high S/N spectra of the A and B binaries. We derive stellar parameters of star A1 from the coadded spectrum. Within the spectroscopic framework iSpec \citep{Blanco-Cuaresma}, we fit synthetic spectra computed using SPECTRUM \citep{Gray_Corbally} and ATLAS9 atmospheres \citep{2004astro.ph..5087C} to the wavelength region 5000-5500\,\AA. The spectroscopically determined parameters are listed in Table \ref{tbl:RVs}. We derive the stellar mass, radius and age by fitting spectroscopic constraints ($T_{\mathrm{eff}}, \log g$, [Fe/H]) to a grid of BaSTI ischrones \citep{Pietrinferni} using the Bayesian Stellar Algorithm \texttt{BASTA} \citep{Silva-Aguirre}, see Table \ref{tbl:RVs}.  

In Fig.~\ref{fig:tracks}, we show the location (with uncertainties) of star A1 in the $\log \, g-T_{\rm eff}$ plane.  Superposed on the plot are evolution tracks for stars of mass 0.9 to $1.5\,\mathrm{M}_\odot$ (mass increases from left to right) in steps of 0.1\,M$_\odot$.  Moreover, the tracks are color coded according to the isochrones of stellar evolution time.

The lines of the primary star in binary B
were too broad ($v\sin i \approx 120$ km/s) to allow for a similar analysis.

\section{Simultaneous Lightcurve and RV-curve Modeling}
\label{sec:model}

\noindent
We carried out a simultaneous analysis of the blended lightcurves of the two eclipsing binaries, as well as the two radial velocity curves of the primaries of the two EBs using our lightcurve emulator code {\tt Lightcurvefactory} \citep{Borko13,Rappaport17}.  This code employs a Markov Chain Monte-Carlo (MCMC)-based parameter search, using our own implementation of the generic Metropolis-Hastings algorithm \citep[see, e.g.][]{Ford}. The basic approach and steps for this study are similar to that which was followed during the previous analysis of the quadruple system EPIC\,220204960, described in \citep[][Sect.\,7]{Rappaport17}. Therefore, here we concentrate mainly on the differences compared to this previous work. 

\subsection{New Features of the Analysis}

First, for a more accurate modeling of the strong ellipsoidal light variation (`ELV') effect in the lightcurve of binary B (see Fig.\,\ref{fig:folds}), we implemented the Roche-equipotential-based stellar surface calculations into our code (see, e.g. \citealp{kopal89}; and \citealp{avni76,wilson79}, for a formal extension to eccentric orbits and asynchronous stellar rotation).  Furthermore, we included an additional switch in the code to set the size parameter of one star (or both) so that it would exactly fill its Roche-lobe. In such a way we were able to model the semi-detached configuration of binary B. 

Second, because our code is now able to fit lightcurve-photometry, radial-velocity, and ETV curves at the same time, we decided to simultaneously analyze the two radial velocity curves along with the blended lightcurve. 

Third, after subtracting off the initial model lightcurves from the data set, we realized that the fluctuations in the residual lightcurve exhibit some distinct periodicities (Fig.\,\ref{fig:zoomLC}, lower panel) with three dominant frequencies which are listed in Table\,\ref{tbl:resfreq}.  We fold the residual lightcurve about the two most significant periods, and plot the two folds separately in the panels of Fig.\,\ref{fig:residual_spots}.  Irrespective of their origin, these variations are modeled in the code in the following automated manner.  In each trial step, after the removal of the blended eclipsing binary model lightcurves from the observed data, the mathematical description of the residual curve is modeled by a harmonic function of the form:
\begin{equation}
\Delta\mathcal{L}=\sum_{i=1}^3a_i\sin(2\pi f_it)+b_i\cos(2\pi f_it),
\end{equation}
where the $f_i$'s are the given, fixed frequencies, and the coefficients $a_i$ and $b_i$ are calculated by a linear least-squares fit. Then, this mathematical model of the residual ligthcurve is added to the binary model lightcurve and the actual $\chi^2$ value is calculated for this mixed model lightcurve. 

\subsection{Significance of the Simultaneous Analysis}
\label{sec:significance}

The main significance of this simultaneous treatment is the following.  Apart from the mass $m_\mathrm{A1}$ and the effective temperature, $T_\mathrm{A1}$, of the primary of binary\,A, all the other astrophysically important parameters of both binaries can be obtained from the same analysis, at least in principle. To prove this statement one needs only recall that both binaries are single-lined spectroscopic binaries (i.e., SB1 systems), and it then follows that the amplitudes of the RV curves give the spectroscopic mass functions 
\begin{equation}
f(m_2)=\frac{(a_1\sin i_1)^3}{P^2}\frac{4\pi^2}{G}=\frac{m_2^3\sin^3i}{(m_1+m_2)^2}=m_1\frac{q^3\sin^3i}{(1+q)^2}.
\end{equation}
Therefore, in the case of binary A, we can use the orbital inclination, $i_\mathrm{A}$, obtained from the blended lightcurve solution, to find the unknown mass $m_\mathrm{A2}$, if we knew $m_\mathrm{A1}$. On the other hand, for binary\,B, which we found to be a semi-detached system, it is expected that its mass ratio, $q_\mathrm{B}$, should be relatively well determined from the lightcurve solution \citep{TerrellWilson05}. Therefore, by combining the spectroscopic mass function, $f(m_2)_\mathrm{B}$, with the mass ratio, $q_\mathrm{B}$, and the inclination angle, $i_\mathrm{B}$, again both obtained from the blended lightcurve solution, one can also calculate the individual masses of the two stars in binary B.

Furthermore, we also wish to point out that the joint photometric analysis of the two binaries inherently carries some information about the mass ratio of the two binaries and well as the temperature ratio of the primary star in each binary ($T_{\rm A1}/T_{\rm B1}$).  Since it turns out that there is already sufficient information to adequately determine all the masses in the system, this means that Eqn.~(\ref{eqn:Tratio}) effectively yields $T_{\rm A1}/T_{\rm B1}$.  Therefore, if $T_{\rm A1}$ is known, one can also find $T_{\rm B1}$ and then, naturally, the effective temperatures of all four stars can also be obtained. Since it is  conceptually interesting that the photometry does encode combined information about the mass ratio of the two binaries and $T_{\rm A1}/T_{\rm B1}$, we provide a brief discussion of this in Appendix\,\ref{app:A}.

\subsection{Fitted Parameters and Assumptions}

As discussed above, all of the astrophysically important parameters of both binaries can be obtained from the same simultaneous analysis, except for the mass $m_{\rm A1}$ and the effective temperature, $T_{\rm A1}$, of the primary of binary A.  However, because $T_\mathrm{A1}$ and its uncertainty are directly known from the spectroscopic analysis, the only remaining task is to find one additional reasonable constraint to close the system of equations. As a good approximation for $m_\mathrm{A1}$ we use the value and uncertainty for $m_\mathrm{A1}$ obtained indirectly from the spectroscopic data, as was described in Sec.\,\ref{sec:NOT-FIES}.

\begin{table}
\centering
\caption{The five most significant peaks of the period analysis of the residual lightcurve.}
\begin{tabular}{lcccc}
\hline
\hline
 & \multicolumn{2}{c}{Frequency} & Amplitude & Phase  \\
 & (d$^{-1}$) & ($\times P_\mathrm{B}$) & ($\times10^{-3}$\,flux) & (rad) \\
\hline
$f_1^a$ & 1.628417(1) & 1.00669 & 7.520(1) & -1.2098 \\
$f_2^a$ & 7.586925(1) & 4.69175 & 3.217(1) & -1.2443 \\
$f_3^a$ & 3.257062(1) & 2.01352 & 2.128(1) & -0.1286 \\
$f_4  $ & 6.102445(1) & 3.77375 & 1.784(1) & -2.7261 \\
$f_5  $ &18.119794(1) &11.20528 & 1.352(1) & -0.8596 \\
\hline
\end{tabular}
\label{tbl:resfreq}

{\bf Notes.} (a) The frequencies used for the lightcurve fitting process.

\end{table}

Turning now to the practical implementation of the combined analysis, we note that in most of the runs we adjusted {20--22} parameters. These are as follows:
\begin{itemize}
\item[(i)]{$2\times3$ orbital parameters: the two periods ($P_\mathrm{A,B}$), inclinations ($i_\mathrm{A,B}$), and reference primary eclipse times ($T_{0,\mathrm{A,B}}$); (Note, in some runs we allowed for an eccentric orbit in binary\,A and, therefore, the eccentricity, $e_\mathrm{A}$, and argument of periastron, $\omega_\mathrm{A}$, of binary\,A were also adjusted, but we did not detect any significant, non-zero eccentricity.  Thus, for most of the runs we simply adopted circular orbits for both binaries.)}
\item[(ii)]{$2\times3$ additional RV-curve related parameters: systemic radial velocities ($\gamma_\mathrm{A,B}$) and linear accelerations ($\dot\gamma_\mathrm{A,B}$)\footnote{Strictly speaking, this latter quantity was taken into account in a slightly unphysical manner; in particular, it was taken to be an absolutely independent variable, and it was not connected to any variation of the eclipsing period.}, and spectroscopic mass-functions ($f(m_2)_{\mathrm{A,B}}$);}
\item[(iii)]{the lightcurve related parameters: temperature ratios $(T_2/T_1)_\mathrm{A,B}$ and also $T_\mathrm{B1}/T_\mathrm{A1}$; the duration of the primary minima $(\Delta t_\mathrm{pri})_\mathrm{A,B}$ \citep[see][Sect. 7 for an explanation]{Rappaport17}; the ratio of stellar radii in binary A $(R_2/R_1)_\mathrm{A}$; and the extra light ($l_\mathrm{x}$);}
\item[(iv)]{the mass ratio ($q_\mathrm{B}$) of binary B;}
\item[(v)] {and finally, the effective temperature $T_\mathrm{A1}$ and mass $m_\mathrm{A1}$ of the primary of binary A, for which we incorporated Gaussian prior distributions with the mean and standard error set to the values obtained from the spectroscopic solution.}
\end{itemize}

Regarding other parameters, a logarithmic limb-darkening law was applied, for which the coefficients were interpolated from the passband-dependent precomputed tables of the {\tt Phoebe} software\footnote{\url{http://phoebe-project.org/1.0}} \citep{Phoebe}. Note, that these tables are based on the stellar atmospheric models of \citet{2004astro.ph..5087C}. The gravity darkening exponents were set to their traditional values appropriate for such late-type stars ($g=0.32$). We found that the illumination/reradiation effect was negligible for the wider binary\,A; therefore, in order to save computing time, it was calculated only for the narrower binary\,B. The Doppler-boosting effect was taken into account for both binaries \citep{LoebGaudi, vanKerkwijk}. 

\begin{figure*}
\begin{center}
\includegraphics[width=0.48 \textwidth]{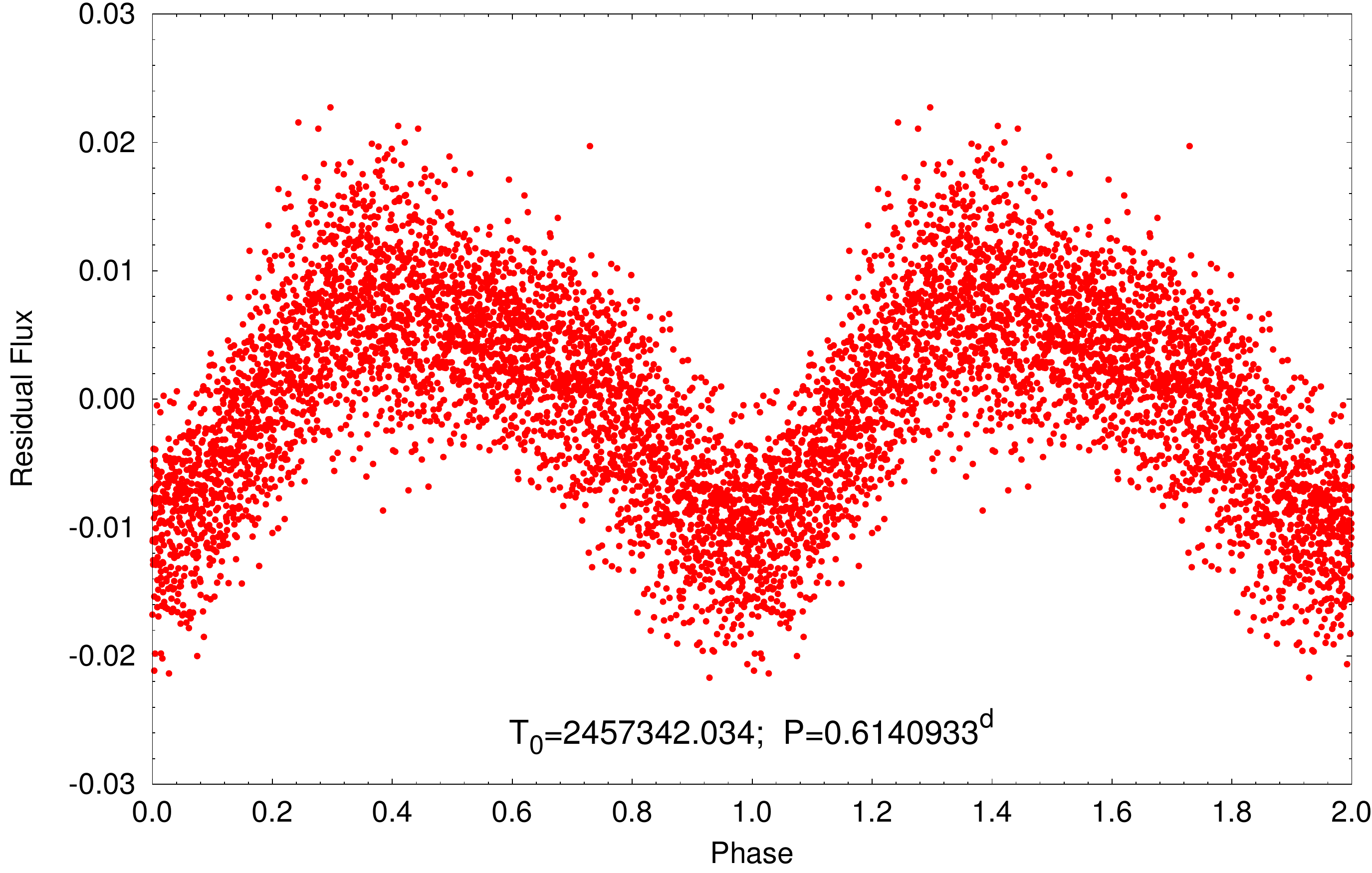}\includegraphics[width=0.48 \textwidth]{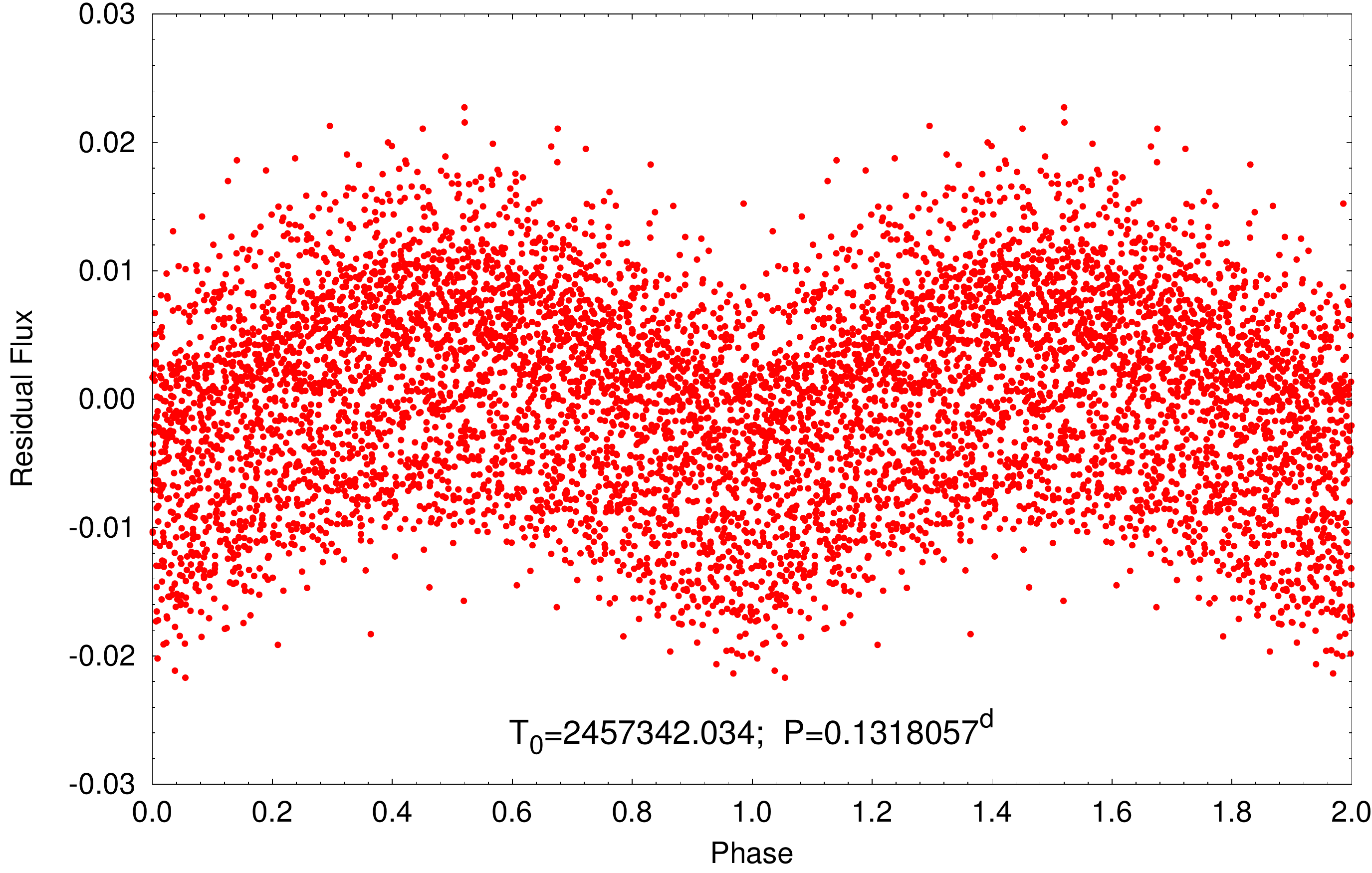}
\caption{Folded lightcurves for periods of 0.6141 and 0.1318 days (left and right, respectively) after the best-fitting orbital lightcurves of the A and B binaries have been subtracted. The period in the left panel is shorter than the orbital period of binary B by only $\sim5.75$ min. We interpret this as starspots from the B binary that are not quite co-rotating with the orbital period of 0.6182 days.}
\label{fig:residual_spots}
\end{center}
\end{figure*}

Furthermore, we assumed that all four stars rotate synchronously with their respective orbits. For the semi-detached component of binary B this assumption seems quite natural. On the other hand, some primaries of semi-detached systems have been found to be rapid rotators relative to their orbits \citep[see, e. g.][for a review]{wilson94}. In our case, however, we may reasonably assume that the highest amplitude peak in the residual lightcurve (see Table\,\ref{tbl:resfreq} and Fig.\,\ref{fig:residual_spots}), with a period which differs by only  $\sim5-6$\,mins from the orbital period of binary B, has its origin in the rotational modulation of the primary of binary B, which clearly dominates the light contribution of this binary. Thus, it is also reasonable to adopt a synchronous rotation for the primary of binary B. Regarding the detached binary A, the spectroscopically obtained projected rotational velocity of the primary component $v\sin i=16.7\pm1\,\mathrm{kms}^{-1}$ (see Table\,\ref{tbl:RVs}) offers an a posteriori verification of our assumption since the projected synchronous rotational velocity that can be deduced from our solution is found to be in essentially perfect agreement with this result (see in Table\,\ref{tbl:simlightcurve}, below). Finally, note that we have no information on the rotation of the secondary component of binary A but, due to its small contribution to the total flux of the system, its rotational properties have only a minor influence on our solution. 

\begin{table*}
\centering
\caption{Parameters from the Double EB simultaneous lightcurve and SB1+SB1 radial velocities solution}
\begin{tabular}{lcccc}
\hline
\hline
Parameter &
\multicolumn{2}{c}{Binary A} & \multicolumn{2}{c}{Binary B}  \\
\hline
$P$ [days]                       & \multicolumn{2}{c}{$3.594728 \pm 0.000014$}   & \multicolumn{2}{c}{$0.618214 \pm 0.000005$} \\
semimajor axis  [$R_\odot$]      &  \multicolumn{2}{c}{$12.21\pm0.19$}           & \multicolumn{2}{c}{$3.65\pm0.16$}  \\  
$i$ [deg]                        & \multicolumn{2}{c}{$89.50\pm0.58$}            & \multicolumn{2}{c}{$64.66\pm0.56$}   \\
$e$                              &  \multicolumn{2}{c}{0}                        & \multicolumn{2}{c}{0} \\  
$\omega$ [deg]                   &  \multicolumn{2}{c}{$-$}                      & \multicolumn{2}{c}{$-$} \\
$t_{\rm prim~eclipse}$ [BJD]     &\multicolumn{2}{c}{$2457341.9183\pm0.0002$}    & \multicolumn{2}{c}{$2457342.0357 \pm 0.0003$} \\
$\gamma$ [km/s]                  & \multicolumn{2}{c}{$30.24\pm0.08$}            & \multicolumn{2}{c}{$27.47 \pm 2.44$} \\  
$\dot\gamma$ [cm/s$^2$]          & \multicolumn{2}{c}{$0.0039\pm0.0004$}         & \multicolumn{2}{c}{$0.0287 \pm 0.0105$} \\  
$f(m_2)$ [$M_\odot$]             & \multicolumn{2}{c}{$0.0863\pm0.0005$}         & \multicolumn{2}{c}{$0.0173\pm0.0027$} \\        
\hline
individual stars & A1 & A2 & B1 & B2 \\
\hline
Relative Quantities: & \\
\hline
mass ratio [$q=m_2/m_1$]         &\multicolumn{2}{c}{$0.56\pm0.05$}     & \multicolumn{2}{c}{$0.31\pm0.03$} \\
fractional radius$^a$ [$R/a$]    & $0.0975\pm0.0014$ & $0.0604\pm0.0014$  & $0.3542\pm0.0077$ & $0.2647\pm0.0077$ \\
fractional luminosity            & $0.37602$          & $0.0186$           & $0.5416$          & $0.0254$ \\
extra light  [$l_\mathrm{x}$]    &\multicolumn{4}{c}{$0.048\pm0.030$} \\
\hline
Physical Quantities: &  \\ 
\hline
$T_{\rm eff}^b$ [K]              & $6473\pm129$  & $4421\pm107$  & $6931\pm250$  & $4163\pm176$ \\
mass$^c$ [$M_\odot$]             & $1.21\pm0.09$ & $0.68\pm0.03$ & $1.30\pm0.21$ & $0.41\pm0.07$ \\
radius$^d$ [$R_\odot$]           & $1.19\pm0.03$ & $0.74\pm0.02$ & $1.33\pm0.06$ & $1.04\pm0.05$ \\
luminosity  [$L_\odot$]          & $2.24\pm0.20$ & $0.19\pm0.02$ & $3.66\pm0.61$ & $0.29\pm0.06$ \\
~        [$M_\mathrm{bol}$]      & $3.87\pm0.10$ & $6.56\pm0.12$ & $3.33\pm0.19$ & $6.09\pm0.22$ \\
$\log \, g$  [cgs]               & $4.37\pm0.04$ & $4.53\pm0.02$ & $4.33\pm0.09$ & $4.10\pm0.09$ \\
$(v\sin i)_\mathrm{sync}^f$ [km/s]& $16.8\pm0.4$ & $10.4\pm0.3$  & $95.6\pm4.8$  & $76.9\pm4.0$  \\
\hline
$(M_V)_\mathrm{tot}$             &\multicolumn{4}{c}{$2.74\pm0.12$} \\
distance$^g$ [pc]                &\multicolumn{4}{c}{$870\pm100$}\\
\hline
\end{tabular}
\label{tbl:simlightcurve}

{\bf Notes.} (a) Polar radii; (b) $T_{\rm eff,A1}$ and its uncertainty were taken from the spectroscopic analysis and used as a Gaussian prior for this joint photometric+RV analysis; the other $T_{\rm eff}$'s were calculated from the adjusted temperature ratios; 
(c) $m_{\rm A1}$ and its uncertainty were taken from the spectroscopic analysis and used as a Gaussian prior; the other masses were calculated as described in Sect.\,\ref{sec:significance}; (d) Stellar radii were derived from the volume-equivalent fractional radii ($R/a$) and the orbital separation; (f) Projected synchronized rotational velocities, calculated using the volume-equivalent radii; (g) Distance to the quadruple, calculated from the photometric distance modulus with the inclusion of an estimate of the interstellar extinction.
\end{table*}

\subsection{Results of the Simultaneous Analysis}

The orbital elements of the two binaries, and the astrophysically relevant parameters of the four stars, together with their uncertainties, are tabulated in Table\,\ref{tbl:simlightcurve}.  About half of these quantities were obtained directly from our simultaneous MCMC analysis of the photometric and RV data, while the others were calculated from the MCMC adjusted parameters using the relations discussed above, as well as some additional trivial ones.  Examples of the latter include the calculation of the semi-major axes from the stellar masses and periods, and the determination of the volume-equivalent physical radii of the four stars from their fractional radii.

We also computed the luminosities of the four stars both in solar luminosity ($L_\odot$) and as bolometric absolute magnitudes. We also compute the total absolute visual magnitude $(M_V)_\mathrm{tot}$ of the quadruple system as a whole. For this latter quantity, the bolometric correction for each star was calculated with the formulae of \citet{flower96}.\footnote{The original coefficients listed in \citet{flower96} contained typos which were corrected by \citet{torres10}. Naturally, these corrected coefficients were used in this work.}  Furthermore, for the calculation of $(M_V)_\mathrm{tot}$ we assumed that the extra light contribution ($l_\mathrm{x}=0.048\pm0.030$) found in our lightcurve solution from the {\em Kepler}-photometric band, is essentially the same as the contaminating light in $V$-band. (However, since it appears that the extra light is fairly negligible, this issue is not very important.) These luminosities and magnitudes are reported in Table \ref{tbl:simlightcurve}.  

Then, by the use of the observed $V$-magnitude, listed in Table\,\ref{tbl:mags}, we can estimate a photometric distance to the system.  We first calculate the maximum hydrogen column density between us and the quadruple, $N_H$, using NASA'S HEASARC on-line tools\footnote{ \url{https://heasarc.gsfc.nasa.gov/docs/tools.html}} and find $N_H \lesssim 1.4 \times 10^{21}$ cm$^{-2}$.  We then used a conversion from $N_H$ to $A_V$ taken from \cite{Guver}: $N_H \simeq 2.2 \times 10^{21} \,A_V$.  This yields an extinction of $A_V \lesssim 0.63$.  We also utilized a web-based applet\footnote{\url{ http://argonaut.skymaps.info/query?}} to estimate $E(B-V) = 0.22 \pm 0.03$ which we translate to $A_V = 0.68 \pm 0.09$.  When we propagate the associated uncertainties in all the involved quantities, we find a distance of $870 \pm 100$ pc; this is also tabulated in the last row of Table \ref{tbl:simlightcurve}\footnote{After this work was completed, the Gaia DR2 \citep{lindegren} were released which provide a distance to EPIC 219217635 of $588 \pm 17$ pc.  This is closer than the photometric distance we estimate of $870 \pm 100$ pc, that is based on a approximate extinction of $A_V \simeq 0.65 \pm 0.09$.  In order for the two distances to be reconciled would require either $A_V \simeq 1.5$ or an unrealistic adjustment of the system $M_V$ that we infer from our joint RV and photometric analysis. Another, possibly more likely explanation would be if the finite separation of two binaries on the sky, i.e., $\lesssim 0.05''$ causes the parallactic distance to be adversely affected \citep{szabados}}.

A comparison of those astrophysical parameters of the {\em primary} star in binary A that were obtained both from the spectroscopic (Table\,\ref{tbl:RVs}), and the combined photometric+RV (Table\,\ref{tbl:simlightcurve}) analysis, shows slight but significant discrepancies. In particular, the radius inferred from the joint photometric+RV analysis ($R_\mathrm{A1}^\mathrm{phot}\simeq1.19\pm0.03\,R_\odot$) is 1.2-$\sigma_\mathrm{spec}$ smaller than the spectroscopically inferred radius ($R_\mathrm{A1}^\mathrm{spec}\simeq1.39^{+0.31}_{-0.17}\,R_\odot$). This leads to an RV+photometric $\log g_{\rm A1}$ that is $0.22 \pm 0.16$ dex higher than that determined from the spectroscopic analysis.  

This slight inconsistency in $\log g_{\rm A1}$ should be considered together with the inferred absolute dimensions of the {\em secondary} component of binary A.  While the effective temperature ($T_\mathrm{A2}\simeq 4400\pm100$\,K) and mass ($m_\mathrm{A2}\simeq0.68\pm0.03\,\mathrm{M}_\odot$) of the secondary are in accord with the main-sequence nature of this star to within the $1-\sigma$ uncertainty, the inferred stellar radius ($R_\mathrm{A2}\simeq0.74\pm0.02\,\mathrm{R}_\odot$) reveals a significantly oversized star for its mass. One might imagine that the inconsistency could readily be resolved assuming that the joint analysis failed to obtain the correct value for the ratio of the relative radii of binary A ($r_2/r_1$), which itself was an adjusted parameter.  However, we note that the eclipses in binary A were found to be total (i.e., in the sense of four contact points) and, therefore, in this case the ratio of the stellar radii are relatively well determined.  This discrepancy did lead us to conduct some further tests, initiating new MCMC runs in which the radius of the secondary of binary A was constrained with the use of the \citet{Tout} mass--radius relation. These runs led to significantly worse fits. In particular, that part of the $\chi^2$ sum which was calculated exclusively from the lightcurve solution was found to be higher by about $\approx8-10\%$ in the case of the solutions using a constrained secondary radius. Thus, using all the presently available information on the quadruple, we are able to resolve these discrepancies, but we believe that they are reasonable given the very different inputs, uncertainties and analyses involved.

Binary B is found to be the more interesting of the two binaries from the perspective of stellar evolution theory.  The stellar components of binary B are: $M_{B1} \simeq 1.33\pm0.21 \, \mathrm{M}_\odot$, $R_{B1} \simeq 1.33\pm0.06 \, \mathrm{R}_\odot$ and $M_{B2} \simeq 0.41\pm0.07 \, \mathrm{M}_\odot$, $R_{B2} \simeq 1.04\pm0.05 \,\mathrm{R}_\odot$.  The larger uncertainties in the masses come mainly from the poorer-quality RV curve, which did not allow for a well-determined spectroscopic mass function.  Note, in particular, the greatly oversized radius of the low-mass secondary star compared to its nominal MS radius.  Thus, we provide a separate discussion of the likely evolutionary scenario for this system in Sect.\,\ref{sec:evolution}.

\section{Constraints on the Quadruple's Outer Orbit}
\label{sec:outerorbit}

We now utilize what we have learned about the A and B binaries from the AO imaging, the RV measurements, and the photometric data to place a couple of significant constraints on the outer orbit of the quadruple system.  There are five principle results that help to constrain the outer orbit: (1) upper limits on the angular separation, $\alpha$, of binary A and binary B; (2) the difference in gamma velocities between the two binaries, $\Delta \gamma \equiv \gamma_A - \gamma_B$; (3) the acceleration of the center of mass of binary A, $\dot \gamma_A$; (4) upper limits on $\dot P_A$ and $\dot P_B$ from the photometric ETV curves; and (5) the inferred masses of all four stars in the binaries. It turns out that the limits on $\dot P$ (item 4) are not significant compared to essentially the same constraint set by $\dot \gamma_A$, and we do not consider this any further.

The specific values of these constraints are as follows: \\
\noindent
$\bullet$  $\alpha < 0.05''$ \\
$\bullet$ $\Delta \gamma \equiv \gamma_A - \gamma_B = -1.4 \pm 2.2$ km s$^{-1}$ \\
$\bullet$ $\dot \gamma_A = 0.0024 \pm 0.0007$ cm s$^{-2}$ \\
$\bullet$ $M_{\rm A} = 2.00 \pm 0.06 \, M_\odot$ and $M_{\rm B} = 1.88 \pm 0.06 \, M_\odot$ \\
We hereafter consider the masses of the A and B binaries to be the same to within their statistical uncertainties.

\begin{figure}
\begin{center}
\includegraphics[width=0.99 \columnwidth]{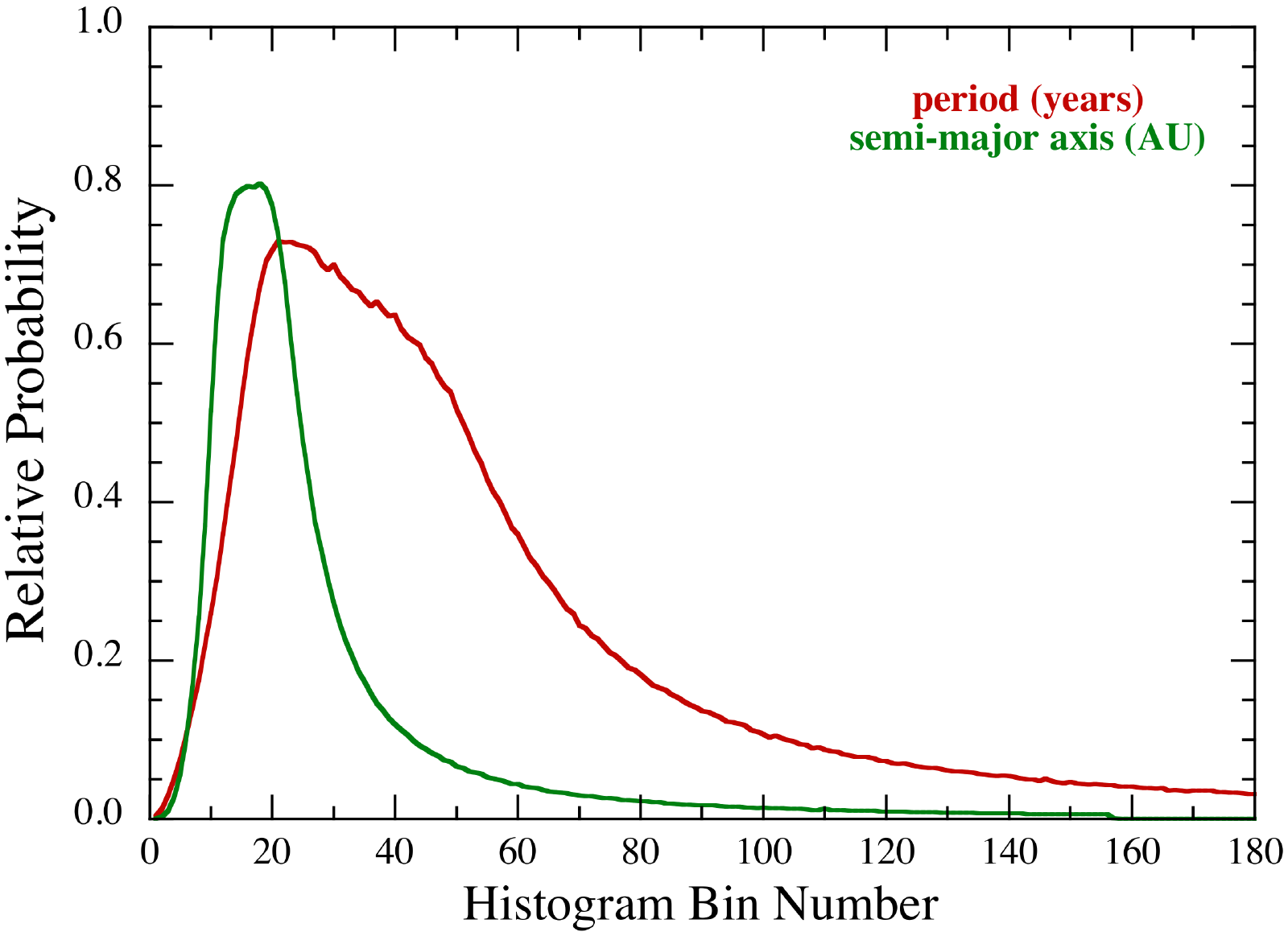}
\caption{Output histograms for the period and semimajor axis associated with the outer orbit of the quadruple.  See text, Sect.~\ref{sec:outerorbit} for details of the Monte Carlo orbit sampling.}
\label{fig:histograms}
\end{center}
\end{figure}  

We now proceed to make use of these facts to constrain the outer orbit.  For an arbitrary outer orbit, we can write down analytic expressions for $\alpha$, $\Delta \gamma$, and $\dot \gamma_A$ (see also \citealt{Lehmann16} and \citealt{Rappaport16}).
\begin{eqnarray}
\alpha  & = & \frac{r}{d} \,  \sqrt{1-\sin^2 i \, \sin^2(\phi+\omega)} ~~~~~\label{eqn:alpha} \\
\Delta \gamma & = & \sqrt{\frac{G M_Q}{a (1-e^2)}} \, \left[\cos(\phi+\omega)+e \cos \omega \right] \sin \, i   \label{eqn:dgamma} \\
\dot \gamma_A  & = & -\frac{G M_B}{r^2}  \, \sin(\phi + \omega) \, \sin \, i  \label{eqn:gammadot}
\end{eqnarray}
The definitions of the quantities appearing in these equations for are: $a$, the semimajor axis; $\phi$, the true anomaly; $\omega$, the argument of periastron; $e$, the orbital eccentricity; and $i$, the orbital inclination angle, where all these quantities pertain explicitly to the outer orbit.  The variable $r$ is the orbital separation, given by the equation of an ellipse: $r = a(1-e^2)/(1+e\cos \phi)$.  Further, $M_Q$ and $M_B$ are the total mass of the quadruple system and binary B, respectively, and $d$ is the distance to the quadruple from the Earth.

There are five parameters of the outer orbit we would like to know ($a$, $P$, $e$, $\omega$, and $i$), and only the four constraints listed above.  Therefore, we will be able to set only ranges of acceptable values for some of these five parameters.  The masses are used to relate $a$ and $P$ through Kepler's third law.

The approach we take to compute probability distributions for $P$, $e$, $\omega$, and $i$ is via Monte Carlo sampling of these parameters, as well as of the unknown instantaneous true anomaly, and then testing for each system realization whether the constraints for $\alpha$, $\Delta \gamma$, and $\dot \gamma_A$ are satisfied to within their uncertainties, assuming Gaussian errors.  For each realization, we randomly sample the mean anomaly in time, compute the corresponding eccentric anomaly, and from that the true anomaly, $\phi$.  Specifically we choose linear random values of $P$ from 0 to 1000 years, $e$ from 0 to 1, 
and $\omega$ from 0 to 2$\pi$.  The orbital inclination was chosen from a uniform probability per unit solid angle.  Finally, the distance to the source was taken to be 870 pc with a Gaussian distribution with $\sigma = 100$ pc. The quantifies $\alpha$, $\Delta \gamma$, and $\dot \gamma_A$ are then evaluated via equations (\ref{eqn:alpha}), (\ref{eqn:dgamma}), and (\ref{eqn:gammadot}) and are compared to the measured values.

Somewhat as we anticipated, the only outer orbit parameters for which interesting constraints could be set are $P$ and $a$.  Output histograms for $P$ and $a$ are shown in Fig.~\ref{fig:histograms}. 
For these two distributions we find 
that the outer period is most probably near 20 years, but could reasonably be as short as 10 years or as long as 80 years.  The corresponding semi-major axis of the quadruple is likely $18 \pm 10$ AU. Therefore, in just a couple of additional seasons of either eclipse monitoring or follow-up RVs from this system, we can expect to see a significant LTTE from the orbit and/or a much more significant determination of $\dot \gamma_A$.

\section{Evolution of Binary B}
\label{sec:evolution}

In this section we address the fact that the lower-mass star in binary B appears to be the more evolved one.  This implies that it is an Algol-like system and the lower-mass star either has lost, or is continuing to lose, its envelope to the currently more massive star.  

Since the properties of the components in binary B are reasonably well determined, this allows us to construct evolutionary scenarios that are self-consistent with the formation and evolution of EPIC 219217635 as a whole. Given the donor star's mass ($0.41 \pm 0.07\, M_\odot$) and radius ($\simeq$ $1.04 \pm 0.05 \,R_\odot$) we conclude that the star is much too large for its mass to be on the main-sequence, and therefore must be substantially evolved.  In fact, we believe that this star belongs to a class of stars known as `Stragglers'  that have been previously studied by \citet{Kaluzny,Orosz03}, and \citet{Mathieu03}. These stars are considerably redder than their main-sequence counterparts and likely experienced some nuclear burning before undergoing a phase of rapid mass-loss (Case AB evolution). Thus Red Stragglers can be legitimately viewed as a special class of Algol-like binaries.

The general properties and evolution of Algol variables have been well-studied (see, e.g., \citealt{Batten}, and references therein; \citealt{Peters}). The more massive star in these systems fills its Roche lobe first and undergoes Roche-lobe overflow (RLOF). A prolonged phase of stable (and sometimes rapid) mass transfer to the less massive companion often ensues.  One of the difficulties in calculating this type of dynamically stable mass transfer arises from the problem of quantifying the degree to which the mass transfer is non-conservative (i.e., to determine the fraction of mass that is lost from the binary during RLOF). As has been shown by \citet{Eggleton2000}, a wide range of values is required in order to explain the observations of Algol-like systems \citep[see, also,][]{NelsonEggleton}.  The secondary usually accretes enough matter so as to cause the mass-ratio to become `inverted' leading to a binary that contains a more evolved yet less massive primary star compared to the secondary star (accretor). Algol variables are normally observed as either being detached with both stars underfilling their respective Roche lobes, or semi-detached with the donor star still undergoing RLOF. The orbital periods of these binaries typically vary from $P_{\rm orb} \gtrsim 1$ day to decades.

Red Stragglers are likely low-mass stars that have evolved considerably (e.g., they may have consumed all of their central hydrogen) before filling their Roche lobes and undergoing a reasonably fast phase of thermal timescale mass transfer to the accretor \citep[see, e.g.,][and references therein]{Zhou18}. The subsequent evolution can be classified with reference to the bifurcation limit \citep{PylyserSavonije}. If mass is stripped rapidly enough, the binaries will evolve below the bifurcation limit and will attain orbital periods on the order of an hour \citep[see, e.g.,][]{Nelson04,Kalomeni16}. In this case the mass-loss timescale of the binary is sufficiently short compared to the donor's nuclear time scale that, although the donor star becomes chemically evolved, it cannot ascend the Red Giant Branch (RGB). On the other hand, if the donor can evolve up the RGB while having its hydrogen-rich envelope stripped away, it will produce a helium white-dwarf remnant \citep[see, e.g.,][]{Rappaport15}. For this latter case, the initial conditions of the progenitor binary allow it to produce a degenerate remnant and thus the binary lies above the bifurcation limit. 

Because the components of binary A were formed coevally with those of binary B and given that the more massive component in A has a mass of $\simeq 1.2 \, M_\odot$ and shows little sign of significant nuclear evolution, we require that the progenitor primary of binary B had a mass of $\gtrsim 1.5 \, M_\odot$. The mass of the progenitor secondary in binary B is much less well-constrained. It must be chosen to be significantly less than that of the primary so that it has not experienced significant nuclear (chemical) evolution and because the mass ratio ($M_{2,0}/M_{1,0}$)\footnote{For purposes of discussing the prior evolutionary history of binary B, we have reversed the labels ``1'' and ``2'', now referring to the originally more massive star as ``1'' and vice versa.} must not be so low as to cause a dynamical instability (leading to a possible merger). We found that progenitor masses of $M_{1,0} \approx 1.7 \, M_\odot$ and  $M_{2,0} \approx  0.8 \, M_\odot$ worked reasonably well in reproducing the currently observed properties of binary B. 

According to our preferred scenario, the progenitor binary consisted of an $\approx 1.7 \, M_\odot$ primary (the current donor star) and a relatively low-mass ($\approx 0.8 \, M_\odot$) secondary. After the primary has burned some of the hydrogen in its core, it undergoes RLOF on its thermal (Kelvin-Helmholtz) timescale. This leads to relatively rapid transfer rates in excess of $10^{-7} M_\odot \, {\rm yr}^{-1}$. After more than one solar mass of material has been lost, the mass of the donor is reduced to the presently inferred value of $\simeq 0.45 \, M_\odot$ while the companion's mass increases to approximately $1.4 \, M_\odot$.  Thus about 50\% of the transferred mass is lost from the binary in the form of a `fast' Jeans wind (the expelled matter carries away the specific angular momentum of the accretor).

In order to test the robustness of the scenario, we created a small grid of evolutionary models using the MESA stellar evolution code \citep{Paxton11}.  The progenitor binary was assumed to have a solar metallicity ($Z=0.02$) and the evolution was computed in accordance with the `standard' RLOF model \citep{GoliaschNelson} under the assumption of a `fast' Jeans mode of systemic mass loss and allowing for gravitational radiation and magnetic braking angular-momentum dissipation. Although highly uncertain, we set the systemic mass-loss parameters such that $\alpha$=0 and $\beta$=0.5 \citep[see][for a detailed explanation]{Tauris06}.  This implies that no mass was ejected from the system directly from the primary, while 50\% of the mass passing through the inner Lagrange point to the secondary was subsequently ejected from the system.  It should be noted that our ability to create models that approximately reproduce the properties of binary B does not depend sensitively on the choice of $\beta$.  We found that adjusting the value of $\beta$ up or down by $\approx 0.2$ would still yield models with similar properties to those of binary B as long as the mass of the progenitor secondary was increased or reduced accordingly.  Although the binary dynamics were computed self-consistently, the changes to the interior structure of each component were calculated independently. The evolutionary tracks were terminated once the secondary (accretor) had evolved sufficiently so as to fill its own Roche lobe.

\begin{figure}
\centerline{\includegraphics[width=0.99\columnwidth]{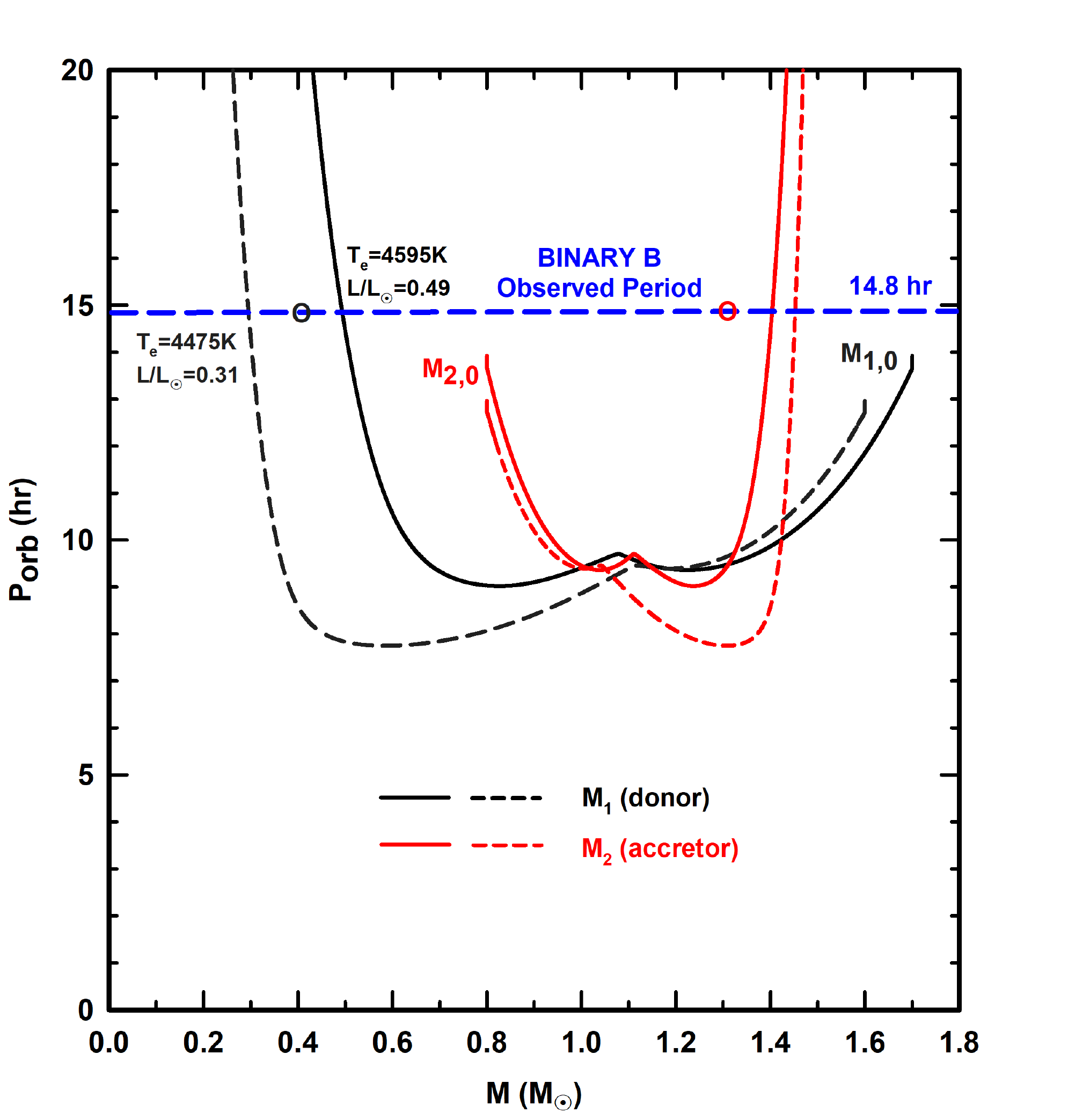}} 
\caption{The evolution of representative models for binary B in the $P_{\rm orb}$ -- $M$ plane. The black and red curves represent the evolution of the primary (donor) and secondary (accretor), respectively. The solid and dashed curves correspond to two different sets of initial conditions for the progenitor binary. The black and red circles denote the locations of the observationally inferred values of $P_{\rm orb}$ and $M$ for the primary and secondary, respectively.  Typical model ages at the current epoch are 2-3 Gyr.}
\label{fig:evolution}
\end{figure} 

Fig.\,\ref{fig:evolution} shows the evolutionary tracks in the $P_{\rm orb}$ -- $M$ plane for two representative sets of initial conditions for the progenitor binary. Starting with $M_{1,0} =  1.7 \, M_\odot$  and $M_{2,0} = 0.8 \, M_\odot$, and an initial $P_{\rm orb} = 14$ hours, the computations imply present-day masses of $0.49 \, M_\odot$ and $1.40 \, M_\odot$, respectively, at a $P_{\rm orb} = 14.8$ hours (solid black curve in Figure 1). The value of $P_{\rm orb}$ is known extremely precisely while the inferred values of the component masses given in Table \ref{tbl:simlightcurve} are less certain. The values of the computed masses are reasonably close to the inferred values for $P_{\rm orb} = 14.8$ hours.  Because of the constraints of Roche geometry imposed on the lobe-filling donor (i.e., the dependence of $R_1$ on $P_{\rm orb}$ and $M_1$), the computed radius agrees with the inferred one to within $\approx 10$\%. We also conclude that the value of $\log g$ is within 0.1 dex of the inferred value. The largest discrepancy can be found in the effective temperature of the donor star.  Our computed $T_{\rm eff}$ is always about 500\,K higher than the inferred value regardless of the initial conditions that we choose{\footnote{ This may be an artefact of the uncertainty in the conversion of colors to temperatures.}}. As an illustration of this point, consider the dashed black track in Figure 1. The initial mass of the primary was chosen to be 1.6 M$_\odot$ and the initial period was $P_{\rm orb} = 13$ hours (the initial secondary mass was the same). According to this track the mass of the present-day donor is reduced to $0.30 \, M_\odot$ but the value of $T_{\rm eff}$ is only reduced by about 100\,K. We could not find combinations of progenitor parameters (or variations in the input physics associated with systemic mass loss) that led to much smaller temperatures. As for the secondary, its computed mass is very close to the inferred value (see the solid red curve in Figure \ref{fig:evolution}). Since the calculated mass transfer rate for the observed orbital period is $\lesssim 10^{-9} M_\odot \, {\rm yr}^{-1}$, the secondary easily relaxes to its approximate thermal equilibrium configuration as it accretes matter. For the present-day, our calculations show that the secondary underfills its Roche lobe by nearly a factor of two. For a mass of $1.4 \, M_\odot$, we find that its temperature is close to 6600\,K and the radius is about $1.6 \, R_\odot$. These values are not significantly different from those given in Table \ref{tbl:simlightcurve}. We also find that of the progenitor models that can reasonably explain the currently observed system properties, evolutionary ages ranged between $\sim $2-3 Gyr.

According to our proposed scenario, the progenitor primary experiences a phase of rapid mass-loss on its Kelvin time while the value of the mass ratio ($M_2/M_1$) is $\lesssim 1$. When the donor's mass is thus reduced to less than $1 \, M_\odot$, it starts to evolve on a nuclear time scale and the radius of the donor increases (as does $P_{\rm orb}$). Although this binary evolves above the bifurcation limit (i.e., the donor would eventually become a giant and collapse to become a helium white dwarf), the increase in the mass of the secondary allows it to evolve and fill its Roche lobe before the donor can become a giant. This might lead to a reversal in the direction of mass transfer or a merger might ensue. Regardless of the possible future evolution, we feel confident in stating that the evolution of binary B can be explained without the need to invoke a phase of common envelope evolution. If this is true, then the study of the evolution of binaries A and B can be carried out independently.

\section{Summary and Conclusions}
\label{sec:conclusion}

In this work we have identified a physically bound quadruple system comprised of two short-period eclipsing binaries in an $\sim$20 AU orbit about each other.  The doubly eclipsing system was found in Field 7 of the K2 mission, with periods of $P_A \simeq 3.5949$ d and $P_B \simeq 0.6182$ d.  

We acquired follow-up ground-based observations including: (1) Keck AO images showing that the separation between the two binaries is $\lesssim 0.05''$; (2) 20 radial velocity measurements with the NOT-FIES spectrometer which yield single-line RV curves for both binaries; and (3) photometry with small-aperture telescopes (12-14 inches) which yielded nine additional eclipse times, thereby increasing the overall observation interval to nearly two years.  

We analyse the photometric and radial velocity data for both binaries all simultaneously to yield many of the binary system parameters.  The results are summarized in Table \ref{tbl:simlightcurve}.  

The eclipse timing variations of both binaries show erratic behavior (binary B) or non-secular trends (binary A), and these are not associated with any light-travel time effects or physical interactions between the binaries.

We set significant constraints on the outer (i.e., quadruple) orbit using the AO and RV measurements.  These indicate that the semi-major axis of the outer orbit is $18 \pm 10$ AU with a likely outer period of 20-40 years   

The upper limit to the angular separation (from the AO image), the nearly matching $\gamma$ velocities, and the similar luminosities of the two binaries provide compelling {\em circumstantial} evidence for the physical association of the two binaries.  By contrast, the detection, at the 3.2-$\sigma$ confidence level, of $\dot \gamma$ for the A binary is important {\em direct} evidence that the binaries are 
physically interacting with one another.  

If a few further RV measurements can be made over the next year, the significance of the $\dot \gamma$ detection can be made much stronger.  As discussed in the Introduction, there are only a relative handful of double eclipsing quadruples known to be physically bound, and EPIC 219217635 is nearly certain to join their ranks.

We have demonstrated that the target is sufficiently bright for small (i.e., 12-14 inch class) telescopes to continue to follow the timing of the primary eclipses of both binaries.  At some point, the light-travel time delays in the system will begin to dominate over the more erratic ETV behavior, and would also allow for a more definitive measure of the outer orbit. For example, in the case of a circular outer orbit having a period of $P_{\rm out}$\,$\approx$\,20 yr, and therefore, an orbital separation of $a_{\rm out}$\,$\approx$\,12\,AU one can expect periodic ETVs for both binaries with almost equal semi-amplitudes of $\mathcal{A}_\mathrm{LTTE}\approx50\times\sin{i_\mathrm{out}}$\,minutes and, of course with opposite phases. Or, from a different perspective, converting the variation of the systemic radial velocity of binary A obtained from our analysis (see Sect.\,\ref{sec:NOT-FIES}), i.e. $\dot\gamma=0.0024\pm0.0007\,\mathrm{cm}\,\mathrm{s}^{-2}$ into a period variation rate, one gets $\Delta P_\mathrm{A}=8.8\pm2.5\times10^{-8}\,\mathrm{day/cycle}$. Assuming that this value is approximately constant over an interval which is much shorter than the outer orbital period, we find that the expected difference of the observed and linearly predicted eclipse times after $N$ inner orbital cycles can be calculated as
\begin{equation}
\Delta t\approx\frac{1}{2}\Delta P\times N^2 \, .
\end{equation}
From this, one can easily show that it is inevitable that there will be an observable 15-minute shift in the eclipse times after only $N\approx477$ cycles, i.e. $\approx4.7$\,years.  Therefore, we can expect definite confirmation of the gravitationally bound nature of this quadruple within a few years.

Note also, that accurate future observations of the complete LTTE orbits of the two binaries will offer all of the benefits which can be obtained from RV measurements of a double-lined spectroscopic binary.  And, in addition, because the masses of the two binaries are known relatively well, one will also be able to calculate from the LTTE amplitudes the observed inclination ($i_\mathrm{out}$) of the outer orbit.

We would also like to suggest that a few additional radial velocity measurements be made for the next few observing seasons for this system. For outer orbital periods of $\sim$20 years, the value of $\dot \gamma_A$ would not only be firmed up, but within just a few more years, the curvature of the outer orbit should be detected.

Finally, from our analysis, B binary appears to have its less massive and cooler star evolved well beyond where its main-sequence radius would be, and is filling (or nearly filling) its Roche lobe. We describe a possible evolutionary path to explain this apparent very short-period Algol-like red-straggler system.

\vspace{0.3cm}

We are grateful to Jules Halpern for acquiring an initial image of the field at the MDM observatory.  
T.\,B. acknowledges the financial support of the Hungarian National Research, Development and Innovation Office -- NKFIH Grant OTKA K-113117. 
S.\,A. and A.\,J. acknowledge support by the Danish Council for Independent Research, through a DFF Sapere Aude Starting Grant nr.\ 4181-00487B.
A.\,V.'s work was supported in part under a contract with the California Institute of Technology (Caltech)/Jet Propulsion Laboratory (JPL) funded by NASA through the Sagan Fellowship Program executed by the NASA Exoplanet Science Institute.
M.\,H.\,K., D.\,L., and T.\,L.\,J.~acknowledge Allan R. Schmitt for making his lightcurve examining software `LcTools' freely available.  
L.\,N.~thanks A. Senhadji for technical assistance and the Natural Sciences and Engineering Research Council (Canada) for financial support provided through a Discovery grant.  We also thank Calcul Qu\'ebec, the Canada Foundation for Innovation (CFI), NanoQu\'{e}bec, RMGA, and the Fonds de recherche du Qu\'{e}bec - Nature et technologies (FRQNT) for computational facilities.  
The radial velocity spectral observations were made with the Nordic Optical Telescope (NOT), operated by the Nordic Optical Telescope Scientific Association at the Observatorio del Roque de los Muchachos, La Palma, Spain, of the Instituto de Astrofisica de Canarias.
The authors are grateful to Davide Gandolfi for time sharing some of his NOT observations between programs.
Some of the data presented in this paper were obtained from the Mikulski Archive for Space Telescopes (MAST). STScI is operated by the Association of Universities for Research in Astronomy, Inc., under NASA contract NAS5-26555. Support for MAST for non-HST data is provided by the NASA Office of Space Science via grant NNX09AF08G and by other grants and contracts. 
A portion of this work was based on observations at the W.~M.~Keck Observatory granted by the California Institute of Technology. We thank the observers who contributed to the measurements reported here and acknowledge the efforts of the Keck Observatory staff. We extend special thanks to those of Hawaiian ancestry on whose sacred mountain of Mauna Kea we are privileged to be guests. 
Some results are based on data from the Carlsberg Meridian Catalogue 15 Data Access Service at CAB (INTA-CSIC). 
The Pan-STARRS1 Surveys (PS1) and the PS1 public science archive have been made possible through contributions by the Institute for Astronomy, the University of Hawaii, the Pan-STARRS Project Office, the Max-Planck Society and its participating institutes, the Max Planck Institute for Astronomy, Heidelberg and the Max Planck Institute for Extraterrestrial Physics, Garching, The Johns Hopkins University, Durham University, the University of Edinburgh, the Queen's University Belfast, the Harvard-Smithsonian Center for Astrophysics, the Las Cumbres Observatory Global Telescope Network Incorporated, the National Central University of Taiwan, the Space Telescope Science Institute, the National Aeronautics and Space Administration under Grant No. NNX08AR22G issued through the Planetary Science Division of the NASA Science Mission Directorate, the National Science Foundation Grant No. AST-1238877, the University of Maryland,  E\"otv\"os L\'or\'and University (ELTE), the Los Alamos National Laboratory, and the Gordon and Betty Moore Foundation.

\onecolumn

\appendix

\section{Mass and Temperature Information from Joint Photometric Solution}
\label{app:A}

Here we show that the depth of the primary eclipse in binary A has encoded in it information about either the mass ratio of the two binaries or the temperature ratio of the primary stars in the two different binary subsystems.  We start by writing down an expression for the depth of the primary eclipse, $D_{A1}$, of binary A, which is a complete transit.  For simplicity, we take the stars to (1) be spherical, and can thereby represent area ratios as, e.g., $(R_{\rm A2}/R_{\rm A1})^2$, and (2) have surface a brightness proportional to $T_{\rm eff}^4$.  However, the derivation would be the same if we did the exercise for stars whose surfaces follow Roche geometry and have their fluxes measured through specific filter bands, and are subject to limb darkening and other higher order effects.  Furthermore, we also assume that any `third-light' contribution, exterior to the quadruple, is negligible. 
\begin{equation}
D_{A1} \simeq \frac{R_{\rm A2}^2 T_{\rm A1}^4}{R_{\rm A1}^2 T_{\rm A1}^4+R_{\rm A2}^2 T_{\rm A2}^4+R_{\rm B1}^2 T_{\rm B1}^4+R_{\rm B2}^2 T_{\rm B2}^4}
\end{equation}
Dividing by $R_{\rm A1}^2 T_{\rm A1}^4$ yields
\begin{equation}
D_{A1} \simeq \frac{(R_{\rm A2}/R_{\rm A1})^2}{1+(R_{\rm A2}/R_{\rm A1})^2 (T_{\rm A2}/T_{\rm A1})^4+(R_{\rm B1}/R_{\rm A1})^2 (T_{\rm B1}/T_{\rm A1})^4+(R_{\rm B2}/R_{\rm A1})^2 (T_{\rm B2}/T_{\rm A1})^4}
\end{equation}
Finally, if we write the scaled radii as lower case ``$r$'', e.g., $r_{\rm A1} \equiv R_{\rm A1}/a_A$, where $a_A$ is the semi-major axis of binary A, then the above expression can be written as
\begin{equation}
D_{A1} \simeq \frac{(r_{\rm A2}/r_{\rm A1})^2}{1+(r_{\rm A2}/r_{\rm A1})^2 (T_{\rm A2}/T_{\rm A1})^4+({\bf a_\mathrm{B}/a_\mathrm{A}})^2(r_\mathrm{B1}/r_\mathrm{A1})^2({\bf T_{\rm B1}/T_{\rm A1}})^4\{1+(r_{\rm B2}/r_{\rm B1})^2 (T_{\rm B2}/T_{\rm B1})^4\}}
\label{eqn:Tratio}
\end{equation}
where all of the terms in this expression are determined directly from the photometric analysis, except for the terms $(a_\mathrm{B}/a_\mathrm{A})$ and $(T_\mathrm{B1}/T_\mathrm{A1})$ in bold face which are the ratio of physical semi-major axes of the two binaries, and the ratio of the effective temperatures of the two primaries.  Thus, in principle, the simultaneous photometric solution of the two binaries contains information on not just radius and temperature ratios, but also on the ratio of semi-major axes, and hence the mass ratio of the two binaries.  


\begin{thebibliography}{}

\bibitem[Ahn et al.(2012)]{Ahn} Ahn, C.P., Alexandroff, R., Prieto, C.A., et al. 2012, ApJS, 203, 21

\bibitem[Auvergne et al.(2009)]{CoRoT} Auvergne, M., Bodin, P., Boisnard, L., et al., 2009, \aap, 506, 411

\bibitem[Avni(1976)]{avni76} Avni, Y. 1976, ApJ, 209, 574

\bibitem[Bagnuolo \& Gies(1991)]{Bagnuolo} Bagnuolo, W.G., Jr., \& Gies, D.R. 1991, ApJ, 376, 266

\bibitem[Balaji et al.(2015)]{balajietal15} Balaji, B., Croll, B., Levine, A. M., \& Rappaport, S. 2015, \mnras, 448, 429

\bibitem[Batten \& Hardie(1965)]{battenhardie65} Batten, A. H., \& Hardie, R. H., 1965, \aj, 70, 666

\bibitem[Batten(1989)]{Batten} Batten, A .H. 1989, {\it Algols; Proceedings of the 107th IAU Colloquium, Sidney, Canada, Aug. 15-19, 1988}, SSRv, 50, 1

\bibitem[Blanco-Cuaresma et al.(2014)]{Blanco-Cuaresma} Blanco-Cuaresma, S., Soubiran, C., Heiter, U., \& Jofr\'e, P. 2014, A\&A, 569, 111

\bibitem[Borkovits et al.(2013)]{Borko13} Borkovits, T., Derekas, A., Kiss, L.~L., Kir\'aly, A., Forg\'acs-Dajka, E., B\'{\i}r\'o, I.~B., Bedding, T.~R., Bryson, S.~T., Huber, D., \& Szab\'o, R., 2013, MNRAS, 428, 1656

\bibitem[Borkovits et al.(2016)]{Borko16} Borkovits, T., Hajdu, T., Sztakovics, J., Rappaport, S., Levine, A., B\'ir\'o, I.B., \& Klagyivik, P. 2016, \mnras, 455, 4136

\bibitem[Borucki et al.(2010)]{Borucki} Borucki, W.J., Koch, D., Basri, G., et al. 2010, Sci, 327, 977

\bibitem[Caga\v{s} \& Pejcha(2012)]{Cagas} Caga\v{s}, P., \& Pejcha, O. 2012, A\&A, 544, L3

\bibitem[Castelli \& Kurucz(2004)]{2004astro.ph..5087C} Castelli F., \& Kurucz R.~L., 2004, astro, arXiv:astro-ph/0405087   

\bibitem[Chambers et al.(2016)]{Chambers} Chambers, KC., Magnier, E.A., Metcalfe, N., et al. 2016, arXiv:1612.05560

\bibitem[Cutri et al.(2013)]{Cutri} Cutri, R.M., Wright, E.L., Conrow, T., et al.~2013, wise.rept, 1C.

\bibitem[De Rosa et al.(2014)]{derosaetal14} De Rosa, R. J., Patience, J., Wilson, P. A., et al., 2014, \mnras, 437, 1216

\bibitem[Eggleton(2000)]{Eggleton2000} Eggleton, P.P. 2000, New AR, 44, 111

\bibitem[Eggleton \& Kiseleva-Eggleton(2001)]{eggletonkiseleva-eggleton01} Eggleton, P. P., \& Kiseleva-Eggleton, L., 2001, \apj, 562, 1012

\bibitem[Erikson et al.(2012)]{eriksonetal12} Erikson, A., Santerne, A., Renner, S., et al., 2012, \aap, 539, A14

\bibitem[Fabrycky \& Tremaine(2007)]{fabryckytremaine07} Fabrycky, D., \& Tremaine, S., 2007, \apj, 669, 1298

\bibitem[Fang et al.(2018)]{fangetal18} Fang, X., Thompson, T. A., Hirata, Ch. M., 2018, \mnras, tmp 494

\bibitem[Fern\'andez Fern\'andez \& Chou(2015)]{fernandezchou15} Fern\'andez Fern\'andez, J., \& Chou, D.-Y., 2015, \pasp, 127, 421

\bibitem[Flewelling et al.(2016)]{Flewelling} Flewelling, H.A., Magnier, E.A., Chambers, K.C., et al. 2016, arXiv:1612.0524

\bibitem[Flower(1996)]{flower96} Flower, P. J., 1996, \apj, 469, 355

\bibitem[Ford(2005)]{Ford} Ford, E.B. 2005, AJ, 129 1706

\bibitem[Frandsen \& Lindberg (1999)]{frandsen} Frandsen, S., \& Lindberg, B. 1999, in Astrophysics with the NOT proc., ed.
H. Karttunen, \& V. Piirola 71

\bibitem[Goliasch \& Nelson(2015)]{GoliaschNelson} Goliasch, J., \& Nelson, L. 2015, ApJ, 809, 80

\bibitem[Gray \& Corbally(1994)]{Gray_Corbally} Gray, R.O., \& Corbally, C.J. 1994, AJ, 107, 742

\bibitem[Guver \& \"Ozel(2009)]{Guver} Guver, T., \& \"Ozel, F. 2009, MNRAS, 400, 2050

\bibitem[Hajdu et al.(2017)]{hajduetal17} Hajdu, T., Borkovits, T., Forg\'acs-Dajka, E., et al., \mnras, 471, 1230

\bibitem[He{\l}miniak et al.(2017)]{helminiaketal17} He{\l}miniak, K. G., Ukita, N., Kambe, E., Koz{\l}owski, S. K., Paw{\l}aszek, R., Maehara, H., Baranec, C., Konacki, M., 2017, \aap, 602, A30

\bibitem[Hong et al.(2018)]{hongetal18} Hong, K., Koo, J.-R., Lee, J. W., et al., 2018, \pasp, 130, 054204

\bibitem[Huber et al.(2016)]{Huber} Huber, D., Bryson, S.T., Haas, M.R., et al. 2016, \apjs, 224, 2

\bibitem[Husser et al.(2013)]{Husser} Husser, T.-O., Wende-von Berg, S., Dreizler, S., Homeier, D., Reiners, A., Barman, T., \& Hauschildt, P. H., 2013, A\&A, 553, 6

\bibitem[Kalomeni et al.(2016)]{Kalomeni16} Kalomeni, B., Nelson, L., Rappaport, S, Molnar, M., Quintin, J., \& Yakut, K. 2016, \apj, 833, 83

\bibitem[Kaluzny(2003)]{Kaluzny} Kaluzny, J. 2003, Acta Astronomica, 53, 51

\bibitem[Kopal(1989)]{kopal89} Kopal, Z. 1989, {\it The Roche Problem and its significance for double-star astronomy.} Astrophysics and Space Science Library, (Kluwer, Dordrecht/Boston/London) 

\bibitem[Kov\'acs et al.(2002)]{Kovacs}  Kov\'acs, G., Zucker, S., \& Mazeh, T. 2002, A\&A, 391, 369

\bibitem[Kozai(1962)]{kozai62} Kozai, Y., 1962, AJ, 67, 591

\bibitem[Lee et al.(2008)]{leeetal08} Lee, C.-U., Kim, S.-L., Lee, J. W., et al., 2008, \mnras, 389, 1630

\bibitem[Lehmann et al.(2012)]{Lehmann12} Lehmann, H., Zechmeister, M., Dreizler, S., Schuh, S., \& Kanzler, R. 2012, A\&A, 541, 105

\bibitem[Lehmann et al.(2016)]{Lehmann16} Lehmann, H., Borkovits, T., Rappaport, S., Ngo, H, Mawet, D., Csizmadia, Sz., Forg\'acs-Dajka, E. 2016, ApJ, 819, 33.

\bibitem[Lindegren et al.(2018)]{lindegren} Lindegren, L., Hernandez, J, Bombrun, A., et al. 2018, arXiv:1804.09366.

\bibitem[Lidov(1962)]{lidov62}  Lidov, M. L., 1962, PlanSS, 9, 719

\bibitem[Loeb \& Gaudi (2003)]{LoebGaudi} Loeb, A., \& Gaudi, B.S. 2003, ApJ, 588, 117

\bibitem[Lohr et al.(2015)]{Lohr} Lohr, M.E., Norton, A.J., Gillen, E., Busuttil, R., Kolb, U.C., Aigrain, S., McQuillan, A., Hodgkin, S.T., \& Gonz\'alez, E. 2015, \aap, 578, A103

\bibitem[Mathieu et al.(2003)]{Mathieu03} Mathieu, R.D., van den Berg, M., Torres, G., Latham, D., Verbunt, F., \& Stassun, K. 2003, \aj, 125, 246

\bibitem[Naoz \& Fabrycky(2014)]{naozfabrycky14} Naoz, S., \& Fabrycky, D. C., 2014, \apj, 793, 137

\bibitem[Nelson et al.(2004)]{Nelson04} Nelson, L. A., Dubeau, E. P., \& MacCannell, K. A.  2004, \apj, 616, 1124

\bibitem[Nelson \& Eggleton(2001)]{NelsonEggleton} Nelson, C., \& Eggleton, P.P 2001, \apj, 552, 664

\bibitem[Ngo et al.(2015)]{Ngo} Ngo, H., Knutson, H. A., Hinkley, S., Crepp, J. R., Bechter, E. B., Batygin, K., Howard, A. W., Johnson, J. A., Morton, T. D., Muirhead, P. S. 2015, \apj, 800, 138

\bibitem[Orosz \& van Kerkwijk (2003)]{Orosz03} Orosz, J.A., \& van Kerkwijk, M.H. 2003, \aap, 397, 237

\bibitem[Paxton et al.(2011)]{Paxton11} Paxton, B., Bildsten, L., Dotter, A., Herwig, F., Lesaffre, P, \& Timmes, F. 2011, \apjs, 192, 3

\bibitem[Pejcha et al.(2013)]{pejchaetal13} Pejcha, O., Antognini, J. M., Shappee, B. J., Thompson, T. A., 2013, \mnras, 435, 943

\bibitem[Perets \& Fabrycky(2009)]{peretsfabrycky09} Perets, H. B., \& Fabrycky, D. C., 2009, \apj, 697, 1048

\bibitem[Peters (2001)]{Peters} Peters, G.J. 2001, {\it The influence of binaries on stellar population studies} (Dordrecht: Kluwer Academic Publishers) ASSL, 264, 79

\bibitem[Pietrinferni et al. (2004)]{Pietrinferni} Pietrinferni, A., Cassisi, S., Salaris, M., \& Castelli, F. 2004, ApJ, 612, 168

\bibitem[Pietrukowicz et al.(2013)]{OGLE} Pietrukowicz, P., Mr\'{o}z, P., Soszy\'{n}ski, I. et al. 2013, Acta Astron, 63, 115.

\bibitem[Pollacco et al.(2006)]{SWASP} Pollacco, D. L., Skillen, I., Collier Cameron, A. et al., 2006, \pasp, 118, 1407

\bibitem[Pr\v{s}a \& Zwitter(2005)]{Phoebe} Pr\v{s}a, A., \& Zwitter, T. 2005, ApJ, 628, 426

\bibitem[Pylyser \& Savonije(1988)]{PylyserSavonije} Pylyser, E., \& Savonije, G.J. 1988, \aap, 191, 57

\bibitem[Rappaport et al.(2015)]{Rappaport15} Rappaport, S., Nelson, L., Levine, A., Sanchis-Ojeda, R., Gandolfi, D., Nowak, G., Palle, E., \& Pr\v{s}a, A. 2015, \apj, 803, 82

\bibitem[Rappaport et al.(2016)]{Rappaport16} Rappaport, S., Lehmann, H., Kalomeni, B., et al. 2016, MNRAS, 462, 1812

\bibitem[Rappaport et al.(2017)]{Rappaport17} Rappaport, S., Vanderburg, A., Borkovits, T., et al. 2017, MNRAS, 467, 2160

\bibitem[Service et al.(2016)]{Service} Service, M., Lu, J.R. , Campbell, R., Sitarski, B.N., Ghez, A.M., Anderson, J. 2016, \pasp, 128, id5004 

\bibitem[Shibahashi \& Kurtz(2012)]{TheThing} Shibahashi, H., \& Kurtz, D.W. 2012, \mnras, 422, 738

\bibitem[Silva Aguirre et al. (2015)]{Silva-Aguirre} Silva Aguirre, V., Davies, G.R., Basu, S., et al. 2015, MNRAS, 452, 2127

\bibitem[Skrutskie et al.(2006)]{Skrutskie} Skrutskie, M.F., Cutri, R.M., Stiening, R., et al. 2006, AJ, 131, 1163.

\bibitem[Smart \& Nicastro(2014)]{Smart} Smart, R.L., \& Nicastro, L. 2014, \aap, 570, 87

\bibitem[Szabados (1997)]{szabados} Szabados, L. 1997, Proceedings of the ESA Symposium `Hipparcos - Venice '97', Italy, ESA SP-402, p. 657.

\bibitem[Tauris \& van den Heuvel(2006)]{Tauris06} Tauris, T.M., \& van den Heuvel, E.P.J. 2006, {\it Compact stellar X-ray sources}, eds., W.H.G. Lewin M. van der Klis, p.623.

\bibitem[Telting et al. (2014)]{telting} Telting, J. H., Avila, G., Buchhave, L., et al.~2014, AN, 335, 41

\bibitem[Terrell \& Wilson(2005)]{TerrellWilson05} Terrell, D., \& Wilson, R.E. 2005, Ap\&SS, 296, 221

\bibitem[Tokovinin(2008)]{tokovinin08} Tokovinin, A. 2008, \mnras, 389, 925

\bibitem[Tokovinin(2014)]{Tokovinin14} Tokovinin, A. 2014, AJ, 147, 87

\bibitem[Tokovinin(2018)]{tokovinin18} Tokovinin, A. 2018, \aj, in press, 2018arXiv180206445

\bibitem[Torres(2010)]{torres10} Torres, G. 2010, \aj, 140, 1158

\bibitem[Torres et al.(2017)]{torresetal17} Torres, G., Sandberg Lacy, C. H., Fekel, F. C.; Wolf, M., Muterspaugh, M. W., 2017, \apj, 846, 115

\bibitem[Tout et al.(1996)]{Tout} Tout, C.A., Pols, O.R., Eggleton, P.P., \& Han, Z. 1996, MNRAS, 281, 257

\bibitem[Tran et al.(2013)]{tranetal13} Tran, K., Levine, A., Rappaport, S., Borkovits, T., Csizmadia, Sz., \& Kalomeni, B. 2013, \apj, 774, 81

\bibitem[Udalski et al.(2008)]{udalskietal08} Tokovinin, A. 2014, AJ, 147, 87

\bibitem[Vanderburg \& Johnson(2014)]{AV} Vanderburg, A., \& Johnson, J.A. 2014, PASP, 126, 948

\bibitem[Vanderburg et al.(2016)]{Vanderburg} Vanderburg, A., Latham, D.W., Buchhave, L.A., et al. 2016, ApJS, 222, 14

\bibitem[van Kerkwijk et al.(2011)]{vanKerkwijk} van Kerkwijk, M.H., Rappaport, S., Breton, R., Justham, S., Podsiadlowski, Ph., \& Han Z. 2010, ApJ, 715, 51

\bibitem[Wilson(1979)]{wilson79}  Wilson, R.E., 1979, ApJ, 234, 1054

\bibitem[Wilson(1994)]{wilson94}  Wilson, R.E., 1994, \pasp, 106, 921

\bibitem[Zacharias et al.(2013)]{UCAC4} Zacharias, N., Finch, C.T., Girard, T.M., Henden, A., Bartlett, J.L., Monet, D.G., \& Zacharias, M.I. 2013, ApJS, 145, 44

\bibitem[Zasche \& Uhlar(2016)]{V994Her} Zasche, P., \& Uhla\v{r}, R. 2016, A\&A, 588, 121

\bibitem[Zhou et al.(2018)]{Zhou18} Zhou, G., Rappaport, S., Nelson, L., et al. 2018, \apj, 854, 109


\end{thebibliography}
\end{document}